\documentclass[lettersize,journal]{IEEEtran}
\usepackage{amsmath,amsfonts,amssymb,amsthm,mathtools}
\usepackage{algorithm,algorithmic}
\usepackage{array}
\usepackage{textcomp}
\usepackage{stfloats}
\usepackage{url}
\usepackage{graphicx}
\usepackage{balance}
\usepackage{cite}
\usepackage{subcaption}
\usepackage{xcolor}
\usepackage{booktabs}
\usepackage{multirow}
\usepackage{makecell}
\usepackage{microtype}
\usepackage[hidelinks]{hyperref}
\usepackage[utf8]{inputenc}
\usepackage[capitalize,noabbrev]{cleveref}
\usepackage{caption}
\theoremstyle{plain}

\theoremstyle{definition}

\theoremstyle{remark}

\begin{document}
\title{LWM-CDE: A Representation Space for Wireless Data Reasoning and Transferability}
\author{Sadjad Alikhani, Akshay Malhotra, Shahab Hamidi-Rad, and Ahmed Alkhateeb%
\thanks{Sadjad Alikhani and Ahmed Alkhateeb are with the School of Electrical, Computer and Energy Engineering, Arizona State University, Tempe, AZ 85281 USA (e-mail: alikhani@asu.edu; alkhateeb@asu.edu). Akshay Malhotra and Shahab Hamidi-Rad are with InterDigital, Inc., USA (e-mail: akshay.malhotra@interdigital.com; shahab.hamidi-rad@interdigital.com).}}
\maketitle

\begin{abstract}
Machine learning deployments in real-world wireless communication tasks face significant generalization challenges due to location and environment-specific signal structure, high diversity in data across different deployments, and limited availability of real-world data. Current approaches for assessing data similarity between training and inference (deployment) distributions, as well as evaluating model transferability, suffer from high computational costs and inconsistent performance, leaving critical model deployment and model life cycle management decisions without a principled foundation. To address this, we introduce a dataset similarity framework built upon the feature space of a pretrained wireless foundation model. Our method, \textbf{LWM-CDE} (Contrastive learning of Dataset Embedding), fine-tunes the dataset embeddings of the foundation model using a combination of contrastive and geometry-shaping losses, creating a structured manifold where distance reliably indicates transferability. Extensive experiments on wireless benchmarks show that LWM-CDE achieves stronger correlation with empirical transfer performance than existing metrics while being more computationally efficient. The learned representation space supports more effective and data-efficient decision-making for tasks like source dataset selection, label-aware augmentation, and budgeted pretraining, demonstrating its broader utility across different wireless communication applications.
\end{abstract}


\section{Introduction}
Machine learning (ML) is being increasingly applied to a wide range of wireless communication tasks, from link-state classification and beamforming to spectrum sensing and interference management~\cite{hu2021deep,wang2019deep}. The wireless communication standardization body, 3GPP, is also extensively exploring the use of ML based methods for multiple applications, including beamforming, to be deployed as part of the upcoming standard~\cite{3gppTR38843}. To ensure robustness across diverse deployments, models are often trained or fine-tuned on multiple datasets collected under different hardware, environments, and propagation conditions~\cite{9789336}. This multi-dataset paradigm, however, introduces a set of critical and recurring decisions for practitioners: when adapting a model to a new scenario, \textbf{which existing dataset should be used as the source domain for transfer?} A closely related challenge is identifying distribution shifts in time-evolving datasets in order to determine when re-adaptation is necessary~\cite{dwivedi2019representationsimilarityanalysisefficient}. In the presence of \textbf{label scarcity}, which auxiliary datasets can be safely added to the training set for joint training or data augmentation without degrading performance~\cite{schäfer2023overcomingdatascarcitybiomedical}? Under \textbf{constrained compute or storage}, \textbf{which subset of datasets} should be selected for pre-training to maximize downstream generalization across unseen conditions~\cite{killamsetty2021glistergeneralizationbaseddata,gururangan2020dontstoppretrainingadapt,yu2024matesmodelawaredataselection}?

In practice, such dataset-level decisions are typically guided by domain intuition or expensive ablation studies. Simple similarity measures computed in the raw channel space are unreliable due to the high dimensionality of wireless signals and strong task-irrelevant variability, causing distances to be dominated by nuisance factors such as power scaling, noise, and hardware artifacts rather than transfer-relevant structure. Low-dimensional projections such as PCA or UMAP can partially mitigate these effects~\cite{mcinnes2020umapuniformmanifoldapproximation}, and recent work has shown that task-aware distances in UMAP spaces can correlate with empirical transfer performance~\cite{morais2026wirelessdatasetsimilaritymeasuring}. However, UMAP-based approaches remain computationally expensive, non-differentiable, and disconnected from the underlying wireless data generation process, limiting their practicality and adaptability.

\begin{figure*}[t]
\centering
\includegraphics[width=\textwidth]{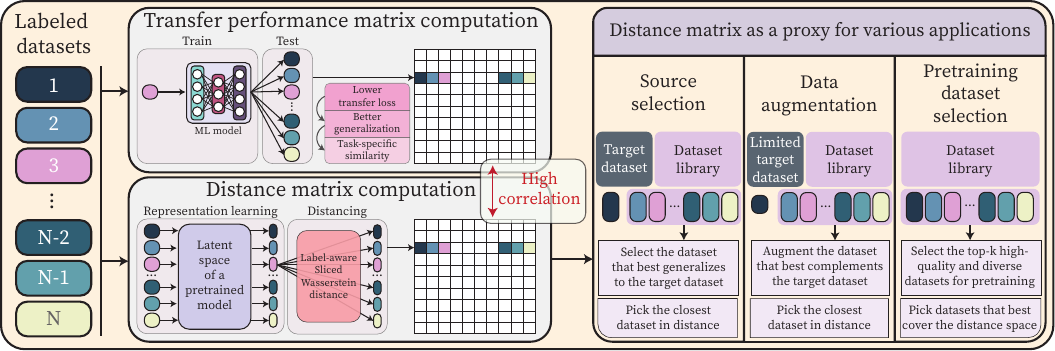}
\caption{\textbf{LWM-based dataset distancing for dataset-level decisions.} From a dataset library, we compute (top) a transfer-performance matrix via source$\rightarrow$target evaluation and (bottom) a distance matrix by embedding channel samples with a pretrained LWM and comparing datasets with label-aware distances. When distances correlate with transfer, the learned geometry becomes actionable and supports source selection, distance-guided augmentation, and budgeted pretraining subset selection.}
\label{fig:lwm_dataset_distancing}
\end{figure*}

\textbf{Foundation-model latent space as a dataset-distancing primitive.} We posit that dataset similarity should be learned in a representation space that is \emph{pre-trained for wireless data} and \emph{fully differentiable}~\cite{mehra2024taskrel,chang2023geometry,shah2024corr,jiang2019fantasticgeneralizationmeasures}. Wireless foundation models such as LWM~\cite{alikhani2024large,11140266} are pre-trained on large diverse wireless corpora and learn semantically meaningful features that linearize a wide range of downstream tasks~\cite{haochen2022separabilityanalyzinglineartransferability}. Used as fixed feature extractors, they already induce representation spaces in which distances between datasets correlate strongly with cross-dataset transfer; crucially, these spaces remain differentiable and can be further refined when transfer supervision is available. The framework is illustrated in \cref{fig:lwm_dataset_distancing}: a transfer matrix obtained by source$\to$target evaluation defines the ground truth for dataset-level decisions, and a distance matrix computed in the LWMs latent space is a faithful and far cheaper surrogate. We use LWM as a frozen encoder, compute label-aware distances (centroid or sliced Wasserstein) in its latent space, and find that even the unrefined geometry substantially outperforms raw-channel baselines and matches UMAP at a fraction of the cost. We then propose \textbf{LWM-CDE} (Contrastive learning of Dataset Embeddings), which fine-tunes a lightweight metric head with a composite contrastive--distillation objective so that distances align more tightly with transfer. We demonstrate utility in three deployment-relevant decision protocols: \emph{source dataset selection}, \emph{distance-guided data augmentation} under scarcity, and \emph{diverse subset selection} for budgeted pre-training.

\textbf{Related work:} Measuring similarity between probability distributions is classical, with theoretical roots in domain-adaptation generalization bounds~\cite{bendavid2010theory,ganin2015dann}. Standard tools include Maximum Mean Discrepancy~\cite{gretton2008kernelmethodtwosampleproblem}, Energy Distance~\cite{SZEKELY20131249}, Centered Kernel Alignment~\cite{kornblith2019similarityneuralnetworkrepresentations}, and optimal-transport distances~\cite{peyré2020computationaloptimaltransport}; sliced Wasserstein~\cite{kolouri2019generalizedslicedwassersteindistances,bonneel2015sliced,liu2025wte} is a tractable approximation we use throughout. For dataset-level reasoning in wireless domains, prior work~\cite{morais2024datasetsimilarityevaluationframework,morais2026wirelessdatasetsimilaritymeasuring} systematically benchmarks these metrics and finds that label-aware Euclidean centroid distance and SWD provide the most consistent correlation with empirical cross-dataset transfer. We adopt these distance functionals and ask a complementary question: in which \emph{representation space} should they be computed? Task embeddings such as Task2Vec~\cite{achille2019task2vec} and Taskonomy~\cite{zamir2018taskonomy} summarise a dataset by a single vector for transferability prediction in vision; our work is the wireless analogue, with the embedding inherited from a domain-pretrained foundation model rather than learned per task. Contrastive frameworks in NLP and vision~\cite{gao2022simcsesimplecontrastivelearning,liu2023rankcseunsupervisedsentencerepresentations,devlin2019bertpretrainingdeepbidirectional,chen2020simclr,he2020moco,caron2020swav,Le_Khac_2020,apple2023betr} use similar listwise and view-consistency objectives at the \emph{sample} level; LWM-CDE adapts the listwise mechanism of RankCSE~\cite{liu2023rankcseunsupervisedsentencerepresentations} to the dataset level and augments it with distribution-to-centroid distillation tailored to wireless dataset comparison.

\textbf{Contributions:} We make three contributions.
\begin{itemize}
\item We identify dataset distancing as a native capability of wireless foundation models: distances in a frozen LWM latent space are transfer-predictive across both held-out DeepMIMO scenarios and the real-world DICHASUS testbed, orders of magnitude faster than UMAP and raw-channel alternatives.
\item We propose \textbf{LWM-CDE}, a lightweight contrastive refinement that fine-tunes a metric head with a four-term objective combining a listwise ranking term, a view-consistency term, a global-correlation term, and a distribution-to-centroid distillation term, and we develop a directed sliced-Wasserstein extension (D-SWD) that handles the asymmetry of the transfer matrix.
\item Across two downstream tasks (LoS/NLoS classification and beam prediction) and three operator-relevant decision protocols, we show the learned geometry approaches the exhaustive oracle within $0.04$ F1 and is robust under label noise, label drop, and full label removal; we also validate it on the real-world DICHASUS testbed under pseudo-LoS/NLoS classification and channel charting.
\end{itemize}

\textbf{Scope:} The framework is designed for \emph{label-aware} transfer prediction: the dataset distances we propose aggregate class-conditional embedding statistics, and their predictive power on downstream transfer is established for tasks in which a per-sample supervisory signal (a discrete label or a continuous regression target such as user position) is available on at least the source side. We do not claim, and do not show, that the same distances are predictive for fully unsupervised transfer problems where the relevant similarity is something like ``noise statistics match''.

\section{Problem Definition}
\label{sec:problem}
Consider a collection $\mathcal{D}=\{D_1,\dots,D_N\}$ associated with a fixed downstream task. For each ordered pair $(i,j)$ we can train a model on $D_i$ and evaluate on $D_j$, yielding a cross-dataset performance score $P_{ij}$. Aggregating these scores forms a generally \emph{directed} transfer matrix
\begin{equation}
\mathbf{P}\in\mathbb{R}^{N\times N}, \qquad P_{ij}\in\mathcal{S},
\label{eq:P_def}
\end{equation}
where $\mathcal{S}$ is the metric range (e.g., $\mathcal{S}=[0,1]$ for normalized error, or for normalized accuracy after a sign flip). When error-type metrics are used, lower values indicate better transfer; for accuracy-type metrics we apply a monotone transformation so that ``better'' corresponds to smaller values. The diagonal terms $P_{ii}$ capture within-dataset performance and serve as a reference for how well the chosen backbone and head perform under no domain shift.

Our objective is to learn a dataset embedding function $\Phi:\mathcal{D}\rightarrow\mathbb{R}^d$ such that geometric proximity predicts transfer behavior. Concretely, we seek a distance $d(\cdot,\cdot)$ for which
\begin{equation}
d\big(\Phi(D_i),\Phi(D_j)\big)\ \text{is predictive of}\ P_{ij},
\label{eq:dist_predicts_transfer}
\end{equation}
up to scaling and monotone transformations. Once such a geometry is learned, it becomes a reusable substrate for dataset-level decisions: selecting a source domain for transfer, identifying distribution shifts in time-evolving deployments, choosing auxiliary datasets for augmentation under label scarcity, and selecting diverse pre-training subsets under compute budgets.

\textbf{Symmetric distances vs.\ asymmetric transfer:} A key modeling choice is that our base distances are \emph{symmetric}, whereas $\mathbf{P}$ is generally \emph{asymmetric}. We evaluate alignment using the symmetrized transfer signal
\begin{equation}
\tilde{P}_{ij}=\tfrac{1}{2}(P_{ij}+P_{ji}),
\end{equation}
and correlate $\{d(D_i,D_j)\}_{i<j}$ with $\{\tilde{P}_{ij}\}_{i<j}$. In our benchmarks, $P_{ij}$ and $P_{ji}$ are often similar, so symmetrization preserves the dominant pairwise structure but cannot capture strongly directional transfer effects. A directed extension that recovers part of the off-diagonal signal is developed in \cref{sec:directed} and evaluated in \cref{sec:analysis_asym}.

\section{Dataset Embeddings and Distance Metrics}
\label{sec:distances}
A fundamental challenge in dataset-level reasoning is identifying a representation space where geometric distances reliably predict functional relationships like cross-dataset transfer. Wireless channel measurements are high-dimensional and complex-valued, with raw-space distances often dominated by task-irrelevant nuisance factors. Consequently, these distances correlate poorly with empirical transferability~\cite{morais2026wirelessdatasetsimilaritymeasuring}. We address this by mapping channel samples into semantically meaningful embedding spaces where nuisance variability is attenuated, then defining dataset-level distances that capture both first-moment mismatches (centroid distance) and full distributional differences (sliced Wasserstein distance).

\subsection{Representation Spaces}
We consider three representation families for mapping channel samples $x\in\mathcal{X}$ to embeddings $z\in\mathbb{R}^d$.

\textbf{Raw channel space:} Channel matrices (or their angle-delay transforms) are vectorized into $\mathbb{R}^{2MS}$, retaining all original variability including nuisance factors.

\textbf{UMAP latent space:} Channels are jointly embedded into a low-dimensional space using UMAP, which preserves local neighborhood structure while suppressing high-dimensional noise. We use standard hyperparameters ($n_{\text{components}}{=}2$, $n_{\text{neighbors}}{=}16$)~\cite{mcinnes2020umapuniformmanifoldapproximation}.

\textbf{Pretrained wireless foundation-model space (LWM):} We use a frozen Large Wireless Model~\cite{alikhani2024large} as a feature extractor. LWM is a Transformer encoder~\cite{vaswani2017attention} pretrained via masked channel modeling on diverse wireless corpora. For a channel $x$ we obtain $\phi_{\text{LWM}}(x)\in\mathbb{R}^{128}$ as either the final-layer \texttt{[CLS]} embedding or mean-pooled token embeddings. The LWM space encodes wireless-domain inductive biases, enables efficient inference via a single forward pass (\cref{sec:cost}), and provides a differentiable mapping that can be refined with task-specific supervision.

\subsection{Dataset-Level Distance Metrics}
Given embedded sample sets $Z_i=\{z_n\}_{n=1}^{m_i}$ and $Z_j=\{z_n\}_{n=1}^{m_j}$ for datasets $D_i$ and $D_j$, we define two complementary distance families.

\textbf{Centroid distance:} The label-aware centroid distance summarizes each dataset by its class-conditional means. For shared classes $\mathcal{C}_{ij}$,
\begin{equation}
d_{\text{CD}}(i,j) = \frac{1}{|\mathcal{C}_{ij}|}\sum_{c\in\mathcal{C}_{ij}}\|\bar{z}_i^{(c)}-\bar{z}_j^{(c)}\|_2,
\end{equation}
where $\bar{z}_i^{(c)}$ is the mean embedding for class $c$ in $D_i$. This metric is efficient but captures only first-moment mismatches.

\textbf{Sliced Wasserstein-2 distance:} To capture higher-order distributional structure~\cite{bonneel2015sliced,kolouri2019generalizedslicedwassersteindistances,peyré2020computationaloptimaltransport}, we approximate optimal transport by averaging 1D Wasserstein distances over random projections. For sampled embedding matrices $Z_i,Z_j\in\mathbb{R}^{n\times d}$,
\begin{equation}
d_{\text{SW}_2}(i,j)\!\approx\!\bigg(\frac{1}{L}\sum_{\ell=1}^{L}\frac{1}{n}\big\|\text{sort}(Z_i u_\ell)-\text{sort}(Z_j u_\ell)\big\|_2^2\bigg)^{\!1/2},
\end{equation}
with $u_\ell\sim\mathcal{U}(\mathbb{S}^{d-1})$. SW$_2$ captures multi-modality and distributional shape but costs more than centroid distance.

\textbf{Label-aware formulation:} For supervised tasks we compute distances only on shared classes to ensure task relevance \cite{morais2026wirelessdatasetsimilaritymeasuring}. When no classes are shared, we fall back to non-label-aware distances on the full sample sets.

\subsection{Why Sliced Wasserstein Predicts Transfer}
\label{sec:sw_proposition}
The empirical correlation between SW distances and cross-dataset transfer is anchored in a classical generalization bound for domain adaptation~\cite{bendavid2010theory}: for hypotheses with bounded Lipschitz loss $\ell$ over the embedding space, the target risk $\varepsilon_T(h)=\mathbb{E}_{(z,y)\sim\mu_T}[\ell(h(z),y)]$ of any classifier trained on source $\mu_S$ is bounded by
\begin{equation}
\varepsilon_T(h)\le\varepsilon_S(h)+L_\ell\cdot W_2(\mu_S,\mu_T)+\lambda^*,
\label{eq:transfer_bound}
\end{equation}
where $W_2$ is the Wasserstein-2 distance, $L_\ell$ the Lipschitz constant, and $\lambda^*$ the joint-optimal error. The label-aware sliced Wasserstein distance we use is a tractable lower bound on $W_2$ between class-conditional distributions on a shared label set: $\mathrm{SW}_2\le W_2$ pointwise on projections~\cite{bonneel2015sliced}, and aggregating per class isolates the component of the bound informative for transfer (label-conditional shifts) from the component the model can absorb via re-weighting (label-marginal shifts). This grounds the use of label-aware $\mathrm{SW}_2$ as a transfer-predictive surrogate; the empirical correlations in \cref{tab:distancing-compact,tab:frozen_vs_cde} quantify how tight the surrogate is in practice.

\section{Contrastive Learning of Dataset Embeddings}
\label{sec:cde}
The frozen LWM space already produces a dataset geometry with informative structure. When a cross-dataset performance matrix $\mathbf{P}$ is available, we further refine this geometry so that distances align more closely with empirical transfer. \textbf{LWM-CDE} is a RankCSE-inspired~\cite{liu2023rankcseunsupervisedsentencerepresentations} procedure that operates at the \emph{dataset} level by fine-tuning a small metric head (and the top LWM layers) so that (i) cosine similarities between dataset centroids respect the row-wise rankings in $\mathbf{P}$ and (ii) label-aware sliced Wasserstein distances correlate with $\mathbf{P}$.

\textbf{Dataset embeddings via stochastic views:} For each $D_i$ we sample $m_i$ examples and pass them through the model twice, with dropout and small Gaussian noise providing stochasticity. This yields two embedding sets $\{z^{(1)}_n\},\{z^{(2)}_n\}$, from which we form $\ell_2$-normalized dataset centroids $u_i=\frac{1}{m_i}\sum z^{(1)}_n$ and $v_i=\frac{1}{m_i}\sum z^{(2)}_n$, treated as two noisy realizations of the same dataset representation.

\textbf{Four-term objective and failure modes:} The training objective combines four terms, each addressing a distinct failure mode observed when fitting the geometry with any single loss in isolation: a \emph{listwise} term aligns the row-wise ordering induced by centroid similarities with the row-wise ordering of $\mathbf{P}$ (without it, the encoder collapses pairwise distances toward a single global scale and loses the per-source ranking that downstream selection consumes); a \emph{consistency} term enforces agreement between two stochastic views of each dataset (without it, the listwise loss becomes brittle under finite-sample fluctuations of the per-row teacher); a \emph{global correlation} term on the upper triangle of the SW distance matrix (without it, the encoder can satisfy every per-row ranking while still distorting the global Pearson correlation that downstream protocols are judged against); and a \emph{distribution-to-centroid distillation} term that propagates the richer SW geometry back to the centroid space (without it, the centroid and SW heads drift apart, and the deployment-time model ends up with weaker neighborhood structure than the SW signal). The first two are adapted from contrastive learning of text embeddings~\cite{liu2023rankcseunsupervisedsentencerepresentations,gao2022simcsesimplecontrastivelearning}; the last two are tailored to dataset-level distributional comparison. \cref{sec:loss_ablation} reports a leave-one-out ablation showing that listwise and distillation are the load-bearing components, with consistency and correlation acting as variance reducers.

\textbf{Listwise performance alignment:} Let $s_{ij}=\langle u_i,u_j\rangle$ denote cosine similarity. For each anchor we define a model distribution $p_{ij}\propto\exp(s_{ij}/\tau)$ and a teacher distribution $q_{ij}\propto\exp(-P_{ij}/\tau)$ over $j\neq i$, so datasets with smaller error $P_{ij}$ receive higher mass. The listwise loss is
\begin{equation}
\mathcal{L}_{\text{list}}=\frac{1}{N}\sum_{i=1}^{N}\text{KL}\big(q_{i\cdot}\|p_{i\cdot}\big).
\label{eq:l_list}
\end{equation}

\textbf{Ranking consistency:} Let $s^{(v)}_{ij}$ and $p^{(v)}_{ij}$ be view-specific similarities and distributions for $v\in\{1,2\}$~\cite{gao2022simcsesimplecontrastivelearning}. The symmetrized KL is
\begin{equation}
\mathcal{L}_{\text{cons}}=\frac{1}{2N}\sum_i\big[\text{KL}(p^{(1)}_{i\cdot}\|p^{(2)}_{i\cdot})+\text{KL}(p^{(2)}_{i\cdot}\|p^{(1)}_{i\cdot})\big].
\end{equation}

\textbf{Global correlation via label-aware sliced Wasserstein:} Let $D^{\text{SW}}_{ij}=d_{\mathrm{SW}}(D_i,D_j)$. We penalize one minus the Pearson correlation between $\mathbf{D}^{\text{SW}}$ and $\mathbf{P}$:
\begin{equation}
\mathcal{L}_{\text{corr}}=1-\rho\!\left(\{D^{\text{SW}}_{ij}\}_{i<j},\{P_{ij}\}_{i<j}\right).
\end{equation}

\textbf{Sliced Wasserstein self-distillation:} The teacher depends only on Wasserstein distances~\cite{hinton2015distillingknowledgeneuralnetwork}, $t_{ij}\propto\exp(-D^{\text{SW}}_{ij}/\tau)$; the student remains $p_{ij}$. The distillation loss is
\begin{equation}
\mathcal{L}_{\text{distill}}=\frac{1}{N}\sum_i\text{KL}(t_{i\cdot}\|p_{i\cdot}).
\end{equation}

\textbf{Overall objective and optimization:} The full LWM-CDE objective is
\begin{equation}
\mathcal{L}_{\text{CDE}}=\lambda_{\text{list}}\mathcal{L}_{\text{list}}+\lambda_{\text{cons}}\mathcal{L}_{\text{cons}}+\lambda_{\text{corr}}\mathcal{L}_{\text{corr}}+\lambda_{\text{distill}}\mathcal{L}_{\text{distill}},
\label{eq:l_cde}
\end{equation}
with $(\lambda_{\text{list}},\lambda_{\text{cons}},\lambda_{\text{corr}},\lambda_{\text{distill}})=(1.0,0.3,0.3,0.3)$ in all experiments. To preserve LWM's pretrained generalization, we fine-tune only a lightweight metric head $g_\phi$ and the top eight Transformer layers, keeping the rest frozen. Optimization uses Adam~\cite{kingma2015adam} with gradient clipping, stochastic dataset-level sampling, and mixed precision. The methodology developed so far is intrinsically \emph{symmetric}: $d(D_i,D_j)=d(D_j,D_i)$. The empirical transfer matrix $\mathbf{P}$ is generally directed, however, and ignoring this asymmetry forfeits an axis of information. The next section develops a directed extension that remains compatible with the rest of the framework.

\section{Directed / Asymmetric Dataset Distances}
\label{sec:directed}
The dataset-level distances of \cref{sec:distances} are symmetric, whereas $\mathbf{P}$ is generally directed: $P_{ij}\neq P_{ji}$ because a balanced, high-quality source typically generalizes better to a skewed target than vice versa. Differences in class balance, sample quality, and within-class diversity all contribute to this asymmetry. We introduce two complementary mechanisms that turn the symmetric label-aware SW distance $d_{\mathrm{SW}}^{(c)}(D_i,D_j)$ (defined per class $c$) into a directed source-to-target estimator $\mathrm{DSW}(D_i\!\to\!D_j)$.

\textbf{Source-prior weighting:} A model trained on source $D_i$ becomes biased toward $D_i$'s dominant classes. Let $\pi_i(c)=|Z_i^{(c)}|/m_i$ be the empirical class prior of source $i$. We weight per-class distances by the source prior to reflect that transfer quality from $i$ to $j$ depends primarily on how well $i$'s dominant classes match $j$'s distribution of the same classes:
\begin{equation}
\mathrm{DSW}_{\pi}(D_i\!\to\!D_j)=\sum_{c\in\mathcal{C}_{ij}}\pi_i(c)\,d_{\mathrm{SW}}^{(c)}(D_i,D_j).
\label{eq:dswd_pi}
\end{equation}
The weights depend on the source identity, making $\mathrm{DSW}_{\pi}(D_i\!\to\!D_j)\neq\mathrm{DSW}_{\pi}(D_j\!\to\!D_i)$ in general.

\textbf{Spread-ratio penalty:} A head trained on tight clusters will fail to generalize to broader target distributions of the same class. Let $\sigma_i^{(c)}$ denote the standard deviation of class $c$ in $D_i$ (e.g., trace-norm of the class-conditional covariance). We add an asymmetric penalty
\begin{equation}
\gamma(i\!\to\!j;c)=\max\!\left(0,\,\log\frac{\sigma_j^{(c)}}{\sigma_i^{(c)}}\right),
\label{eq:spread_pen}
\end{equation}
which fires when the target distribution is broader than the source's. The final directed distance combines \eqref{eq:dswd_pi} and \eqref{eq:spread_pen}:
\begin{equation}
\mathrm{DSW}(D_i\!\to\!D_j)=\sum_{c\in\mathcal{C}_{ij}}\pi_i(c)\,\big[d_{\mathrm{SW}}^{(c)}(D_i,D_j)+\alpha\,\gamma(i\!\to\!j;c)\big],
\label{eq:dswd_full}
\end{equation}
with $\alpha\geq 0$ trading off mean-mismatch and spread-mismatch. We default to $\alpha=1$.

\textbf{Compatibility with LWM-CDE:} The listwise loss in \eqref{eq:l_list} already operates per row on the asymmetric $\mathbf{P}$, so the refined encoder naturally preserves directional information during contrastive training; only the \emph{evaluation distance} needs to be replaced by \eqref{eq:dswd_full}. No re-training is required. The empirical behavior of $\mathrm{DSW}$, including its comparison to symmetric distances against the same unsymmetrized $\mathbf{P}$, its sensitivity to the spread-ratio coefficient, and its degradation under label corruption, is reported in \cref{sec:analysis_asym}.

\section{Evaluation Protocols and Metrics}
\label{sec:protocols}
We evaluate the proposed framework through a controlled empirical study answering two questions: (i) whether distances computed in a learned dataset embedding space reflect empirical cross-dataset transfer, and (ii) whether this geometry can guide concrete, data-efficient decisions in practice.

\subsection{Tasks}
\label{sec:tasks}
We use two supervised tasks that are standard in wireless systems but can be interpreted as generic ML problems with distinct structure.

\textbf{LoS/NLoS classification:} Each example corresponds to a wireless channel observation, and the objective is to predict the propagation condition~\cite{alikhani2024large}. The label indicates whether the signal propagates through a direct path or is obstructed; our setting extends this to multiple propagation categories with pronounced class imbalance across datasets. Labels are obtained from scenario metadata $\eta_n$ via a deterministic rule $y_n=h_{\mathrm{LoS}}(\eta_n)\in\{1,\dots,C\}$, with $C=2$ for binary LoS/NLoS and larger $C$ in our multi-category setting.

\textbf{Beam prediction:} Communication at millimeter-wave frequencies relies on selecting a discrete transmission beam from a finite codebook $\mathcal{F}=\{\mathbf{f}_1,\dots,\mathbf{f}_B\}$~\cite{alikhani2024large,9121328}. The task predicts the optimal beam index from a sub-6\,GHz observation; labels are defined by codebook search on the corresponding mmWave channel $\mathbf{h}_n^{\mathrm{mmW}}$:
\begin{equation}
b_n=\arg\max_{b\in\{1,\dots,B\}}\big|(\mathbf{h}_n^{\mathrm{mmW}})^{\!H}\mathbf{f}_b\big|^2.
\label{eq:beam_label_rule}
\end{equation}
This is a \emph{high-cardinality multi-class} problem with strong cross-dataset variation and limited label transferability, particularly sensitive to the choice of training data.

These tasks span complementary regimes: a classification setting sensitive to label imbalance (LoS/NLoS) and a higher-cardinality task with stronger cross-dataset variability (beam prediction). For each task we construct dataset embeddings using raw channel features, UMAP projections, frozen pretrained wireless representations, and contrastively refined representations.

\subsection{Datasets}
\label{sec:datasets}
All evaluations use 70 ray-traced wireless channel datasets spanning diverse sites worldwide, generated with DeepMIMO~\cite{alkhateeb2019deepmimogenericdeeplearning}, which provides physically grounded channels and rich metadata. Each dataset corresponds to a distinct propagation environment (different geometry, materials, blockage, layout), leading to systematic shift in both channel statistics and labels. \textbf{Channel samples:} each is $x=\mathbf{H}\in\mathbb{C}^{M\times S}$ with $(M,S)=(32,32)$. \textbf{Splits and scale:} each dataset has 3000 training and 2000 validation samples. We use 33 datasets for contrastive fine-tuning of LWM-CDE and hold out the remaining 37 as unseen test datasets to evaluate generalization. Held-out datasets are never used during fine-tuning. \textbf{Evaluation usage:} on the held-out datasets we compute pairwise distances under frozen LWM, LWM-CDE, supervised UMAP, unsupervised UMAP, and raw channel space, and compare against random selection where appropriate.

\subsection{Global Alignment Metrics}
We first assess whether geometric distances align with empirical transfer. For a given embedding space, we compute a distance matrix $D$ and a performance matrix $\mathbf{P}$, flatten the upper triangles, and compute Pearson and Spearman correlation coefficients. Pearson captures approximate linear relationships; Spearman measures rank consistency and is robust to monotone distortions.

\subsection{Decision-Oriented Evaluation Scenarios}
\label{sec:decision_scenarios}
While global correlation is necessary, it is not sufficient for practical utility. We examine three deployment-relevant scenarios.

Let $d(D_i,D_j)\in\mathbb{R}_{\ge 0}$ be the dataset distance induced by a chosen representation and metric. For a downstream task, let $\mathcal{A}(D_s\!\to\!D_t)$ denote empirical transfer performance and $\mathcal{L}(D_s\!\to\!D_t)=1-\mathcal{A}(D_s\!\to\!D_t)$ the transfer loss.

\textbf{Source dataset selection:} When a practitioner wants to deploy a model in a new environment $D_t$ but lacks the budget to label or train from scratch, the cost-effective path is to fine-tune a model trained on the most similar source already in the library. The operational question is which $D_s$ to use, and the standard answer in practice is either a guess based on superficial metadata (city, frequency, antenna count) or an exhaustive ablation across every candidate, infeasible at operator scale. Given a target $D_t$, the oracle source is
\begin{equation}
s^\star(t)=\arg\min_{s\neq t}\mathcal{L}(D_s\!\to\!D_t),
\label{eq:source_oracle}
\end{equation}
which requires the entire row $\mathbf{P}[:,t]$ of the transfer matrix. The distance rule replaces this with a purely geometric query in the embedding space,
\begin{equation}
\hat{s}(t)=\arg\min_{s\neq t}d(D_s,D_t),
\label{eq:source_distance}
\end{equation}
which costs $\mathcal{O}(N)$ distance evaluations after a one-time embedding pass and requires no transfer-matrix evaluations on the target side. The top-$K$ generalization $\hat{\mathcal{S}}_K(t)=\operatorname{TopK}_{s\neq t}\{-d(D_s,D_t)\}$ expands recall at linear cost. We evaluate by comparing achieved transfer $\mathcal{A}(D_{\hat{s}(t)}\!\to\!D_t)$ against the oracle and against random source selection.

\textbf{Distance-guided data augmentation:} Label scarcity is the rule rather than the exception: the labeled portion of an operator-grade library is concentrated in a handful of well-instrumented sites, while the long tail has channel data but no per-sample labels. The natural recourse is to augment the small labeled target $D_t$ with an auxiliary $D_a$, but the choice matters: too distant an auxiliary harms (negative transfer), too similar adds little, and the right operating point sits in between. Picking $D_a$ by exhaustively training augmented variants is expensive precisely where the practitioner has no labels to spare. For target $D_t$ with $n$ labeled samples and auxiliary $D_a$, the augmentation gain is
\begin{equation}
\Delta(a,t;n)=\mathcal{A}_{\mathrm{aug}}(a,t;n)-\mathcal{A}_{\mathrm{base}}(t;n),
\label{eq:aug_gain}
\end{equation}
where $\mathcal{A}_{\mathrm{aug}}$ trains on $D_t^{(n)}\cup\mathrm{Aug}(D_a)$ and $\mathcal{A}_{\mathrm{base}}$ on $D_t^{(n)}$ alone, with $\mathrm{Aug}(D_a)$ a fixed-size sample (a fixed number of samples per class in our protocol). The oracle auxiliary maximizes $\Delta(a,t;n)$; the distance rule selects
\begin{equation}
\hat{a}(t)=\arg\min_{a\neq t}d(D_a,D_t),
\label{eq:aug_distance_select}
\end{equation}
or the top-$K$ nearest auxiliaries for stacked augmentation. The hypothesis is that gain correlates with $-d(D_a,D_t)$: closer auxiliaries supply more transferable structure. We test by measuring rank correlation between $\Delta(a,t;n)$ and $-d(D_a,D_t)$ across candidates, and by comparing distance-guided gains to random auxiliary selection at the same budget.

\textbf{Pre-training subset selection:} Operator-grade libraries grow into the hundreds of datasets, and pretraining on the union is infeasible under realistic compute, storage, and bandwidth budgets. Under a budget of $k\ll N$ datasets, which subset should be pretrained on so that the resulting representation generalizes to held-out targets? The structure differs from source selection: redundant sources give diminishing returns, so a good subset must balance \emph{coverage} of the library with \emph{diversity} among selected datasets. Pure-diversity objectives such as max-sum dispersion select extremes that may not represent the bulk; pure-coverage objectives collapse onto representative centers but leave outliers uncovered. Given a budget $k$, we select $\mathcal{S}\subseteq\mathcal{D}$ with $|\mathcal{S}|=k$ to generalize to a held-out target set $\mathcal{D}_{\mathrm{target}}=\{T_1,\dots,T_M\}$ disjoint from the source pool, scored by
\begin{equation}
\mathcal{G}(\mathcal{S})=\frac{1}{M}\sum_{m=1}^{M}\mathcal{A}\big(\mathcal{E}(\mathcal{P}(\mathcal{S}),T_m)\big),
\label{eq:gen_score}
\end{equation}
with $\mathcal{P}(\mathcal{S})$ pretraining on the union and $\mathcal{E}(\cdot,T_m)$ downstream adaptation. Because $\mathcal{G}$ is expensive combinatorially, we replace it with a distance-only $k$-medoids covering objective:
\begin{equation}
\min_{|\mathcal{S}|=k}\,\sum_{i=1}^{N}\min_{D_j\in\mathcal{S}}d(D_i,D_j).
\label{eq:kmedoids}
\end{equation}
The inner term assigns each dataset to its closest medoid; the outer sum aggregates across the full pool. Minimizing \eqref{eq:kmedoids} forces medoids to be both representative and diverse. We solve it with FasterPAM~\cite{schubert2021fasterpam}. For $k=1$ the objective reduces to the global medoid $j^\star=\arg\min_j\sum_i d(D_i,D_j)$. We compare to random selection and to an empirical upper bound (exhaustive at small $k$; best of $15$ random subsets otherwise).

\section{Experimental Results}
\label{sec:results}
This section answers the central empirical question: do distances in the LWM (and LWM-CDE) latent space predict empirical cross-dataset transfer, and do they support deployment-grade dataset-level decisions? We organize the results around three protocols: source selection and augmentation are reported jointly in \cref{tab:distancing-compact}; end-to-end source-selection payoff in \cref{tab:payoff}; budgeted subset selection in \cref{fig:pretrain,tab:oracle_tm,fig:oracle_ub}; the section closes with real-world DICHASUS validation (\cref{tab:dichasus}). An analysis of why the framework works, when CDE helps, and where it breaks is deferred to \cref{sec:analysis}.

\begin{table*}[t]
\centering
\caption{Correlations for source selection and data augmentation across tasks and embedding spaces.}
\label{tab:distancing-compact}
\renewcommand{\arraystretch}{1.05}
\setlength{\tabcolsep}{12pt}
\begin{tabular}{llcccc}
\toprule
\textbf{Task} & \textbf{Space} & \makecell{\textbf{Source Sel.}\\\textbf{Pearson}} & \makecell{\textbf{Source Sel.}\\\textbf{Spearman}} & \makecell{\textbf{Data Aug.}\\\textbf{Pearson}} & \makecell{\textbf{Data Aug.}\\\textbf{Spearman}} \\
\midrule
\multirow{5}{*}{\makecell{LoS/NLoS\\Classification}}
 & LWM        & $0.772\pm 0.047$ & $0.681\pm 0.042$ & $0.617\pm 0.191$ & \underline{$0.728\pm 0.225$} \\
 & LWM-CDE    & \underline{$0.783\pm 0.115$} & \textbf{$0.778\pm 0.043$} & \textbf{$0.682\pm 0.254$} & \textbf{$0.744\pm 0.214$} \\
 & UMAP-unsup & $0.429\pm 0.132$ & $0.481\pm 0.116$ & $0.484\pm 0.224$ & $0.454\pm 0.341$ \\
 & UMAP-sup   & \textbf{$0.795\pm 0.060$} & \underline{$0.726\pm 0.076$} & \underline{$0.657\pm 0.200$} & $0.602\pm 0.308$ \\
 & Channel    & $0.224\pm 0.113$ & $0.134\pm 0.147$ & $0.329\pm 0.211$ & $0.229\pm 0.201$ \\
\midrule
\multirow{5}{*}{\makecell{Beam\\Prediction}}
 & LWM        & \underline{$0.774\pm 0.096$} & \underline{$0.777\pm 0.086$} & \underline{$0.745\pm 0.229$} & \underline{$0.742\pm 0.225$} \\
 & LWM-CDE    & \textbf{$0.782\pm 0.114$} & \textbf{$0.788\pm 0.103$} & \textbf{$0.775\pm 0.233$} & \textbf{$0.792\pm 0.181$} \\
 & UMAP-unsup & $0.610\pm 0.112$ & $0.618\pm 0.099$ & $0.540\pm 0.338$ & $0.513\pm 0.332$ \\
 & UMAP-sup   & $0.656\pm 0.129$ & $0.634\pm 0.129$ & $0.380\pm 0.183$ & $0.512\pm 0.167$ \\
 & Channel    & $0.586\pm 0.154$ & $0.611\pm 0.140$ & $0.596\pm 0.293$ & $0.600\pm 0.179$ \\
\bottomrule
\end{tabular}
\end{table*}

\subsection{Dataset Distance for Transfer and Augmentation}
\label{sec:transfer_aug}
\cref{tab:distancing-compact} reports Pearson and Spearman correlations between inter-dataset distances and downstream performance across two tasks and five embedding spaces. Three patterns stand out.

\textbf{LWM-based spaces outperform UMAP and raw-channel baselines across both tasks and decision types.} LWM-CDE achieves the strongest and most stable correlations overall, with Spearman for source selection reaching $0.788$ on beam prediction and $0.778$ on LoS/NLoS. For the higher-variance augmentation scenario, LWM-CDE remains robust (Spearman up to $0.792$). Datasets closer in the LWM-CDE space tend to yield better transfer and more reliable augmentation gains, enabling principled rank-based decisions without exhaustive training.

\textbf{The advantage is largest where raw geometry fails.} For LoS/NLoS source selection, raw-channel Pearson is $0.224$ and Spearman $0.134$, while LWM-CDE achieves $0.783$ and $0.778$. This is the regime where a semantically rich, domain-specific representation is necessary: the binary task is dominated by per-environment LoS ratio, which raw distances are blind to. Supervised UMAP performs competitively for source selection but is less consistent for beam prediction and degrades for augmentation. Embeddings optimized solely to preserve label structure miss other forms of variability (spatial layout, propagation, hardware) that matter for transfer; LWM-CDE captures a more general transfer-relevant geometry.

\textbf{Choice of baselines:} We benchmark against raw-channel and UMAP-based distances rather than MMD~\cite{gretton2008kernelmethodtwosampleproblem} or CKA~\cite{kornblith2019similarityneuralnetworkrepresentations} as distance functions because the controlled wireless-domain comparison of these alternatives has been performed in~\cite{morais2024datasetsimilarityevaluationframework,morais2026wirelessdatasetsimilaritymeasuring}, which finds label-aware Euclidean centroid distance and SWD dominate MMD/CKA on cross-dataset transfer prediction. The contribution of this paper is not to re-litigate the choice of distance functional but to show that the \emph{representation space} in which an already-validated distance is computed materially changes its predictive power.

\textbf{Reading the standard deviations:} The width of the uncertainty bands in \cref{tab:distancing-compact} is itself diagnostic. On source selection the LWM-CDE Spearman standard deviation drops to $0.043$ on LoS/NLoS, the lowest of any non-supervised-UMAP row, while raw-channel rises to $0.147$ on the same task. LWM-CDE not only ranks sources better on average but ranks them \emph{more consistently across splits}, the property an operator needs from a selector run once per deployment: a high mean correlation with high variance gives statistically indistinguishable picks on adjacent random subsets, whereas tight bands mean the same source is reliably nominated each time. On augmentation, bands widen across all spaces (LWM-CDE Spearman $\pm 0.214$ on LoS/NLoS) because the augmentation gain $\Delta(a,t;n)$ in \eqref{eq:aug_gain} subtracts two noisy quantities of comparable magnitude and inherits the variance of both, a known limitation of any difference-of-accuracies signal rather than a property of the distance.

\subsection{End-to-End Source-Selection Payoff}
\label{sec:payoff}
Correlations characterize how well dataset distances rank potential sources, but the operator-relevant question is whether using these rankings to actually pick a source yields a measurable downstream gain. We evaluate the full selection-then-train pipeline on a deliberately diverse 24-dataset LoS/NLoS pool drawn from DeepMIMO ($8$ cities $\times$ $3$ BS, LoS fractions in $[7\%,79\%]$) with per-dataset budget $n=400$. For each held-out target $t$ we pick $\hat{s}(t)$ under four strategies: random ($20$ trials), frozen-LWM distance, LWM-CDE distance (with the encoder fine-tuned once on the same pool's $\mathbf{P}$), and a per-target oracle $\arg\min_{s\neq t}P[s,t]$. F1 is read off the transfer matrix.

\begin{table}[t]
\centering
\caption{End-to-end source-selection payoff on the 24-dataset LoS/NLoS pool. F1 averaged over $24$ targets; ``recovered'' is the fraction of the random$\to$oracle gap closed by the distance rule.}
\label{tab:payoff}
\setlength{\tabcolsep}{6pt}\renewcommand{\arraystretch}{1.05}
\begin{tabular}{lcc}
\toprule
Strategy & Mean F1 & \% gap recovered \\
\midrule
Random source                    & $0.773$ & --                     \\
Frozen-LWM nearest source        & $0.861$ & $47.1\%$                 \\
LWM-CDE nearest source           & $\mathbf{0.875}$ & $\mathbf{54.8\%}$ \\
Per-target oracle ($\arg\min P$) & $0.959$ & $100\%$                  \\
\bottomrule
\end{tabular}
\end{table}

\cref{tab:payoff} shows three findings. \emph{First}, random selection averages $F_1=0.773$, the per-target oracle reaches $0.959$, leaving a recoverable gap of $+0.186$. \emph{Second}, frozen-LWM distance closes $47.1\%$ of the gap; LWM-CDE closes $54.8\%$, a $+1.4$-point gain over the strong frozen baseline and $+10.2$ points over random. \emph{Third}, the remaining $\sim 45\%$ of the gap is irreducible by any top-1 distance-based selector with a single-shot pick: it requires either retrieving the per-target argmin of $\mathbf{P}$ exactly (which requires the matrix) or a multi-source selection (the budgeted-pretraining setting below). The correlations in \cref{tab:distancing-compact} therefore translate to a tangible deployment gain.

\subsection{Pretraining Dataset Subset Selection}
\label{sec:subset}
We next study \emph{budgeted pretraining dataset selection}: choosing $k$ source datasets for pretraining under resource constraints while maximizing generalization to held-out targets. Redundant sources yield diminishing returns, so this setting requires reasoning about diversity and coverage. We select sources using $k$-medoids over the pairwise distance matrix. The candidate pool is $N=24$ sub-6\,GHz datasets; the held-out target pool is $8$ disjoint cities ($1$ BS each); $n_{\text{train}}=n_{\text{test}}=400$. For each subset of size $k$, one head is trained on the union of the $k$ sources and evaluated on each held-out target. We compare: (i) random ($5$ seeds), (ii) UMAP centroid distance + $k$-medoids ($3$ seeds), (iii) frozen-LWM SW + $k$-medoids ($3$ seeds), (iv) LWM-CDE SW + $k$-medoids ($3$ seeds), with the encoder fine-tuned once on the $24$-pool $\mathbf{P}$, and (v) an empirical oracle (every singleton at $k=1$; best of $15$ random subsets otherwise).

\begin{figure}[t]
\centering
\begin{subfigure}[b]{0.48\linewidth}\centering
\includegraphics[width=\linewidth]{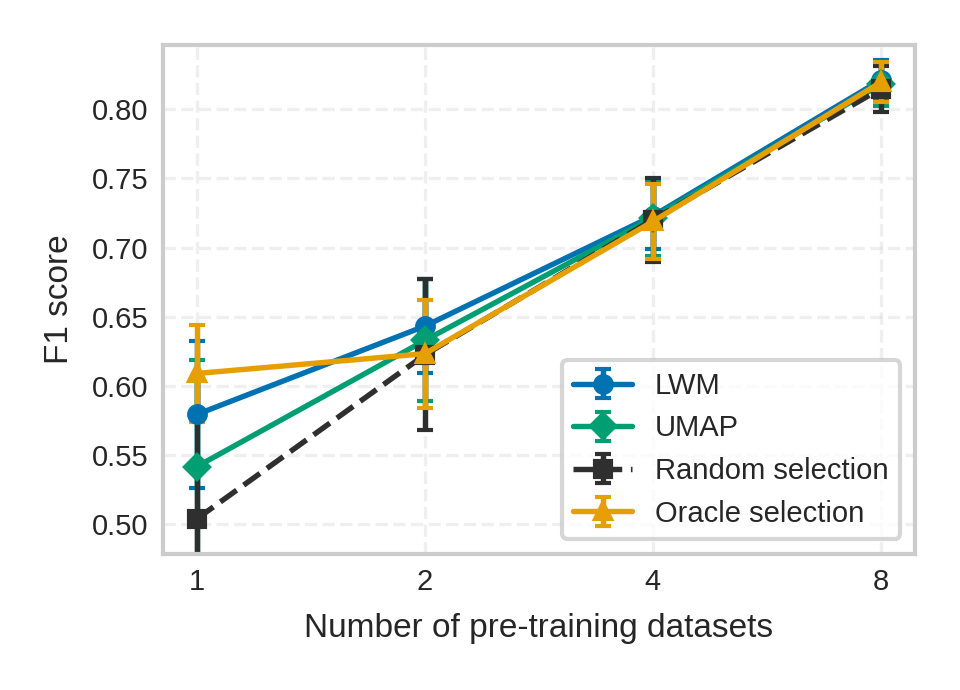}
\caption{Beam prediction}\label{fig:pretrain_bp}\end{subfigure}\hfill
\begin{subfigure}[b]{0.48\linewidth}\centering
\includegraphics[width=\linewidth]{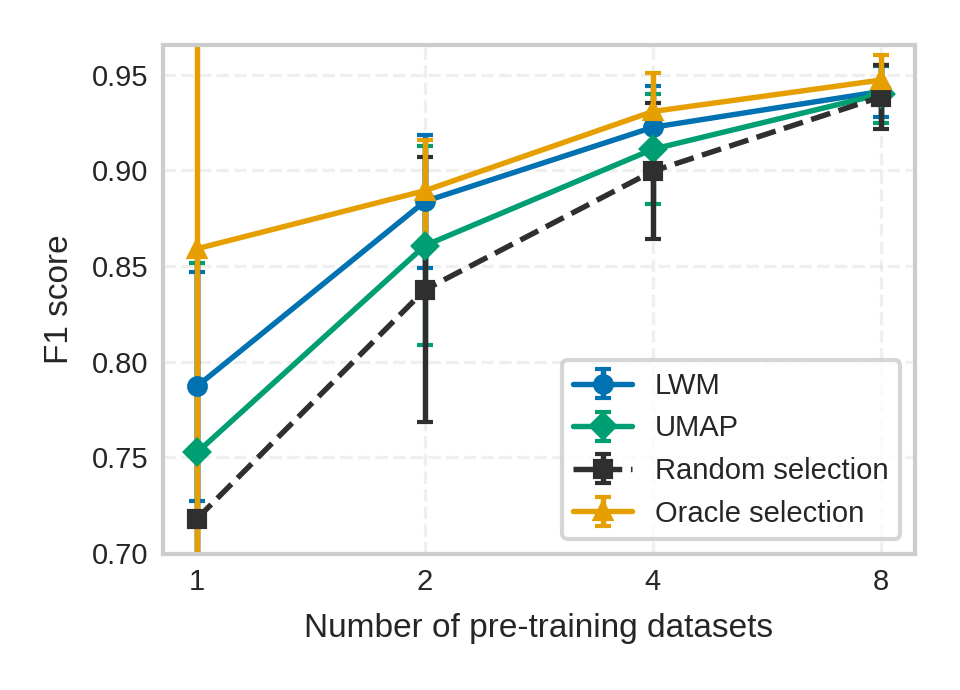}
\caption{LoS/NLoS classification}\label{fig:pretrain_los}\end{subfigure}
\caption{Pretraining subset selection under a dataset budget $k$. For each curve, $k$ source datasets are chosen from a 24-dataset diverse sub-6 GHz pool ($8$ cities $\times 3$ base stations, mixing high-LoS US and low-LoS international sites) using $k$-medoids over the indicated distance matrix; one classifier head is then trained on the union of those $k$ sources and evaluated on $8$ held-out target cities. Mean F1 across the $8$ targets is reported, with error bars over $3$ selection-and-training seeds for the distance-based methods, $5$ seeds for random, and $15$ random subsets for the empirical oracle.}
\label{fig:pretrain}
\end{figure}

\textbf{At low budgets, distance-guided selection materially beats random.} On LoS/NLoS at $k=1$ (\cref{fig:pretrain_los}), random averages $F_1=0.73$ with standard deviation $0.30$: its worst trial collapses to $F_1=0.14$ when the picked source is Shanghai ($\text{LoS}\approx 7\%$) and targets are mostly high-LoS US cities, so the head over-predicts NLoS. Distance-based methods avoid this pathology by selecting medoids that cover the high-LoS regime: UMAP $0.86$, frozen LWM $0.88$, LWM-CDE $0.91$, against an empirical oracle of $0.97$. The ordering holds at $k=2$ and narrows by $k=4$. \textbf{On beam prediction the picture is similar at $k\geq 2$:} frozen-LWM-medoids reaches $F_1=0.90$ at $k=2$ vs $0.85$ for random and $0.87$ for UMAP and LWM-CDE; all methods converge to the oracle $0.97$ by $k=8$. At $k=1$ the four methods are within $0.005$ of each other ($\approx 0.81$) because $k$-medoids picks the same global medoid from each distance matrix; the $0.92$ gap to the oracle at this budget reflects the fundamental limit of any single-source selector for a multi-class task with target-specific best sources.

\textbf{All methods saturate near the empirical oracle by $k=8$.} The distance-based selectors close $80$--$90\%$ of the random-to-oracle gap in the low-$k$ regime where each chosen dataset is most consequential, and LWM/LWM-CDE remain within $0.005$ F1 of the oracle at $k\geq 4$.

\begin{figure}[t]
\centering
\begin{subfigure}[b]{0.48\linewidth}\centering
\includegraphics[width=\linewidth]{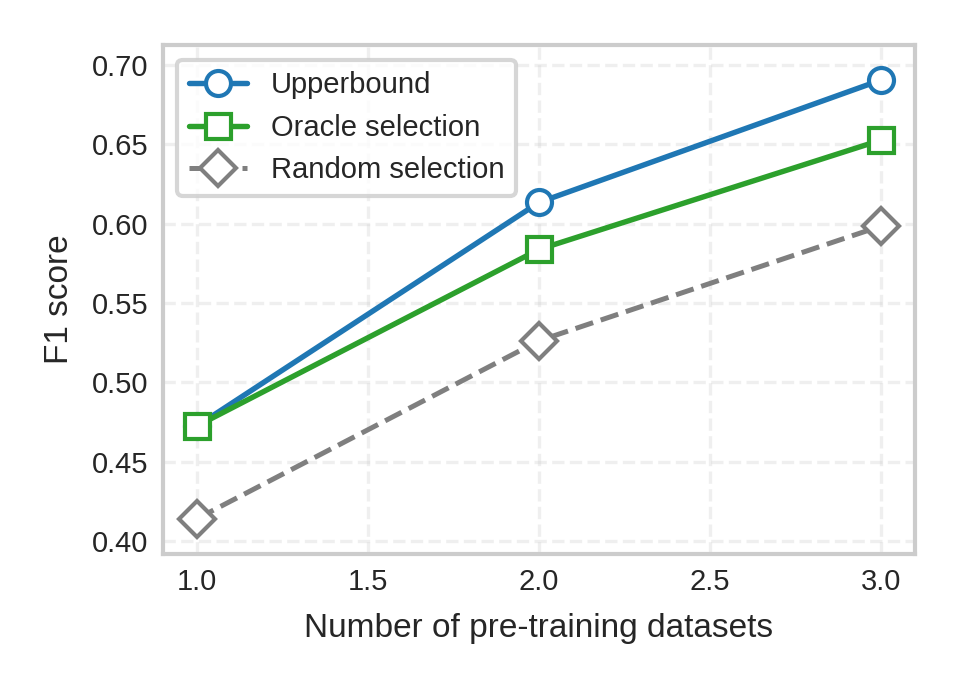}
\caption{Beam prediction}\label{fig:oracle_ub_bp}\end{subfigure}\hfill
\begin{subfigure}[b]{0.48\linewidth}\centering
\includegraphics[width=\linewidth]{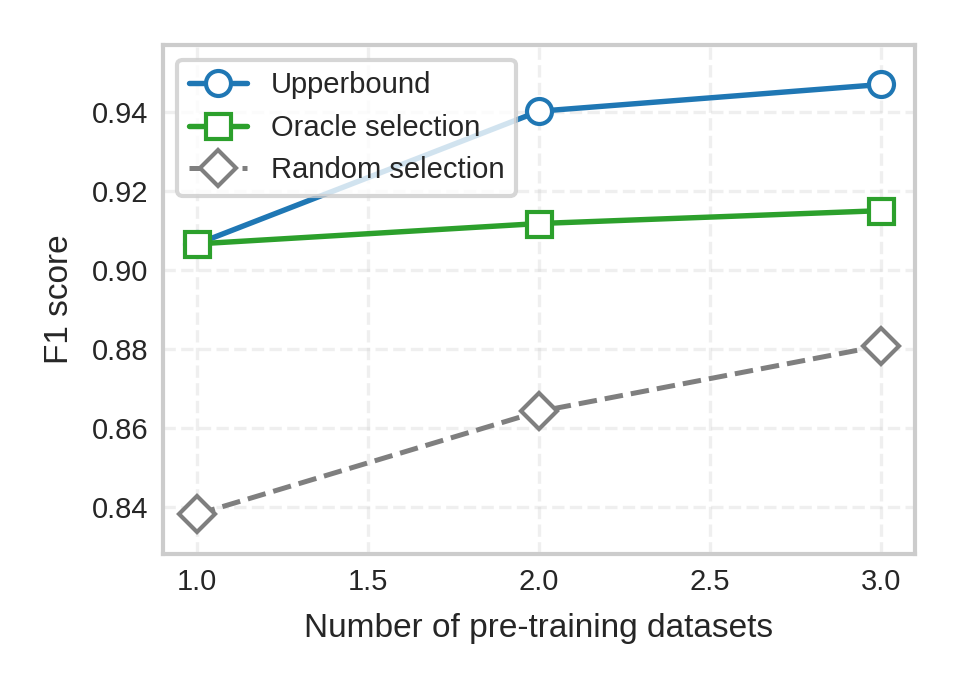}
\caption{LoS/NLoS classification}\label{fig:oracle_ub_los}\end{subfigure}
\caption{True upper bound (UB) vs.\ transfer-matrix oracle for task-specific dataset selection. For each task and $k\in\{1,2,3\}$, we compare the best-performing subset obtained by exhaustive evaluation (UB) with subsets selected by TM-$k$-medoids based on the directed transfer-performance matrix.}
\label{fig:oracle_ub}
\end{figure}

\begin{table}[t]
\centering
\caption{Regret and ranking agreement of transfer-matrix $k$-medoids (TM). Regret is the gap between true UB and TM-$k$-medoids. Kendall's $\tau$ measures agreement between the subset ranking induced by TM-$k$-medoids and the true ranking.}
\label{tab:oracle_tm}
\setlength{\tabcolsep}{5pt}
\begin{tabular}{ccccc}
\toprule
& \multicolumn{2}{c}{\textbf{LoS/NLoS}} & \multicolumn{2}{c}{\textbf{Beam Prediction}} \\
\cmidrule(lr){2-3}\cmidrule(lr){4-5}
$k$ & Regret$\downarrow$ & Kendall $\tau\uparrow$ & Regret$\downarrow$ & Kendall $\tau\uparrow$ \\
\midrule
1 & $0.0000$ & $0.7333$ & $0.0000$ & $0.6856$ \\
2 & $0.0284$ & $0.7149$ & $0.0296$ & $0.6303$ \\
3 & $0.0319$ & $0.5864$ & $0.0378$ & $0.4529$ \\
\bottomrule
\end{tabular}
\end{table}

\textbf{Oracle validation against exhaustive search.} \cref{fig:oracle_ub} compares the exhaustive UB, the TM oracle, and random. The TM oracle (diversity-aware: best singleton at $k=1$, $k$-medoids over $\mathbf{P}$ at $k\geq 2$) closely tracks the exhaustive UB and consistently beats random. \cref{tab:oracle_tm} quantifies this: TM-$k$-medoids attains near-zero regret at $k=1$ and remains within $0.04$ F1 for $k\in\{2,3\}$, with Kendall's $\tau\geq 0.45$. This validates distance-based selection as a near-oracle proxy when exhaustive evaluation is infeasible.

\subsection{Real-World Validation on the DICHASUS Testbed}
\label{sec:dichasus}
A central concern for AI-native 6G is whether geometries learned on synthetic data generalize to real measurements. We adopt the \textbf{DICHASUS} indoor $1.272$\,GHz campaign ($4127$ samples, $4$ antennas, $1024$ subcarriers, tachymeter-verified positions). The LWM encoder is \emph{not} re-trained on DICHASUS; the same encoder used on synthetic data is used here.

\textbf{Multi-dataset testbed:} Because DICHASUS is a single campaign, we synthesize a benchmark by partitioning receivers into $Z=7$ spatial zones via $k$-means on $(x,y)$ and treating each zone as an independent dataset. Different zones exhibit distinct multipath, blockage, and LoS conditions, so cross-zone transfer is non-trivial.

\textbf{Pseudo-LoS/NLoS classification:} Per-sample received power is thresholded at the median to create binary labels; for each zone pair $(i,j)$ we train a lightweight MLP on $i$ and report $P_{ij}=1-F_1$ on $j$, averaged over $10$ trials.

\textbf{Channel charting (position regression):} We also evaluate an unsupervised-style task: predicting normalized $(x,y)$ positions from raw channel patches~\cite{studer2018channelcharting,ferrand2021channelcharting}. Here $P_{ij}$ is cross-zone NMSE, averaged over $5$ trials. This tests whether dataset distances reflect similarity in the channel-to-position mapping itself, independent of any classification head.

\begin{table}[t]
\centering
\caption{Real-world validation on DICHASUS: global distance--transfer correlations on cross-zone transfer. The LWM encoder is \emph{not} re-trained on DICHASUS.}
\label{tab:dichasus}
\setlength{\tabcolsep}{6pt}\renewcommand{\arraystretch}{1.05}
\begin{tabular}{llcc}
\toprule
Task & Embedding & Pearson & Spearman \\
\midrule
\multirow{2}{*}{Pseudo-LoS/NLoS}
   & \textbf{LWM}  & \textbf{$0.64\pm 0.08$} & \textbf{$0.54\pm 0.09$} \\
   & UMAP          & $-0.01\pm 0.13$         & $-0.02\pm 0.21$ \\
\midrule
Channel charting & \textbf{LWM} & \textbf{$0.74\pm 0.04$} & \textbf{$0.66\pm 0.04$} \\
\bottomrule
\end{tabular}
\end{table}

\cref{tab:dichasus} reports global distance--transfer correlations. On both tasks, distances in the frozen LWM space yield strongly positive correlations, while UMAP distances collapse to near-zero. \textbf{Despite being trained exclusively on synthetic ray-traced channels, the LWM space carries enough propagation-relevant structure to remain a meaningful surrogate for transfer on out-of-distribution real measurements.} This is the property required for AI-native 6G: a foundation model trained once on broad simulated data can be reused as a dataset-distancing primitive in deployment scenarios it never saw at training time. The drop in absolute correlation strength reflects two factors we deliberately do not control: per-zone sample counts are $5$--$10\times$ smaller than DeepMIMO splits, so per-class SW estimates are noisier; and pseudo-labels from received-power thresholding are noisier than DeepMIMO's geometry-derived labels. The fact that LWM-space distances retain a meaningful signal under both effects, while UMAP does not, is the central message.

\section{Analysis}
\label{sec:analysis}
The previous section established that distances in the LWM (and LWM-CDE) latent space predict empirical transfer well enough to support concrete dataset-level decisions. This section probes the mechanisms. We ask in turn: when is contrastive refinement worth its one-time cost (\cref{sec:frozen_vs_cde})? Which loss components drive the alignment (\cref{sec:loss_ablation})? Which distance metric is the right choice (\cref{sec:euc_vs_swd})? How much directional information is recoverable (\cref{sec:analysis_asym})? How robust is the framework to label corruption (\cref{sec:robustness})? Does refinement transfer across bands (\cref{sec:xband})? What is the computational cost (\cref{sec:cost})?

\subsection{When Does Contrastive Refinement Help?}
\label{sec:frozen_vs_cde}
We compute global correlations of label-aware SW distances against $\tilde{\mathbf{P}}$ on the held-out test split with identical sample budgets ($n=1000$ per dataset, $L=64$ projections, per-class cap $M=200$); the CDE encoder is fine-tuned once on the 33-dataset training split via \eqref{eq:l_cde} and not retuned at test time.

\begin{table}[t]
\centering
\caption{Frozen LWM vs.\ LWM-CDE on global distance--transfer correlations (symmetric label-aware SW, held-out split).}
\label{tab:frozen_vs_cde}
\setlength{\tabcolsep}{6pt}\renewcommand{\arraystretch}{1.05}
\begin{tabular}{lcccc}
\toprule
& \multicolumn{2}{c}{Frozen LWM} & \multicolumn{2}{c}{LWM-CDE} \\
\cmidrule(lr){2-3}\cmidrule(lr){4-5}
& Pearson & Spearman & Pearson & Spearman \\
\midrule
Beam pred.\ (source sel.) & $0.85$ & $0.75$ & \textbf{$0.86$} & \textbf{$0.76$} \\
LoS/NLoS (source sel.)    & $0.64$ & $0.68$ & \textbf{$0.78$} & \textbf{$0.78$} \\
Beam pred.\ (data aug.)   & $0.745$ & $0.728$ & \textbf{$0.775$} & \textbf{$0.792$} \\
LoS/NLoS (data aug.)      & $0.617$ & $0.728$ & \textbf{$0.682$} & \textbf{$0.744$} \\
\bottomrule
\end{tabular}
\end{table}

\textbf{The frozen LWM space is already a strong dataset embedder.} Without any task-specific supervision the frozen encoder reaches Pearson $0.85$ on beam prediction and $0.64$ on LoS/NLoS, both well above raw-channel and unsupervised UMAP baselines. This validates wireless foundation models as out-of-the-box dataset embedders for deployments where CDE refinement is infeasible.

\textbf{The CDE gain is task-geometry-dependent.} On beam prediction the gap is small ($+0.01$ to $+0.05$): the high-cardinality problem is already nearly linearizable in the frozen LWM space, and the residual ordering errors that CDE corrects are small. On LoS/NLoS the gap is large ($+0.14$ Pearson on source selection, $+0.10$ Spearman): the task is binary, dominated by per-environment LoS ratio ($7\%$--$79\%$), and the listwise loss in \eqref{eq:l_list} directly penalizes ordering violations of the form ``a more LoS-heavy source generalizes better to a balanced target than to an NLoS-heavy target'', a structure the frozen encoder does not encode unaided.

\textbf{The qualitative ranking is robust to sample budget.} Re-running LoS/NLoS source selection with $n=400$ across $10$ random splits, frozen LWM gives Pearson $0.589\pm 0.113$ / Spearman $0.612\pm 0.083$ and CDE gives $0.637\pm 0.117$ / $0.652\pm 0.116$. Both numbers are below the large-budget values, which is expected: the variance of any $U$-statistic estimator decays as $\mathcal{O}(1/n)$, so smaller per-class samples raise the noise floor. The CDE refinement still yields a consistent positive gap, and the relative ordering of methods is preserved, which is what dataset-level decision protocols actually require.

\textbf{Fine-tuning is amortized across every subsequent decision.} Forty epochs over the 33-dataset training set complete in roughly ten minutes on a single GPU, comparable to a single source-target evaluation in the exhaustive ablation it replaces; once trained, the encoder is fixed and reused.

\subsection{Loss Component Ablation}
\label{sec:loss_ablation}
\cref{tab:contrastive_ablation_bp} switches the four LWM-CDE loss terms on and off and measures Pearson/Spearman against the transfer error matrix $E=1-\mathbf{P}$ on beam prediction.

\begin{table}[t]
\centering
\caption{\textbf{Contrastive loss ablation (Beam Prediction).}}
\label{tab:contrastive_ablation_bp}
\setlength{\tabcolsep}{5pt}\renewcommand{\arraystretch}{1.05}
\footnotesize
\begin{tabular}{cccccc}
\toprule
\multicolumn{4}{c}{\textbf{Loss terms}} & \textbf{Pearson} & \textbf{Spearman} \\
\cmidrule(lr){1-4}
List & Cons & Corr & Dist & & \\
\midrule
 &  &  &  & $0.725\pm 0.186$ & $0.653\pm 0.250$ \\
\midrule
\checkmark &  &  &  & $0.777\pm 0.130$ & $0.709\pm 0.196$ \\
 & \checkmark &  &  & $0.759\pm 0.132$ & $0.694\pm 0.196$ \\
 &  & \checkmark &  & $0.762\pm 0.133$ & $0.692\pm 0.239$ \\
 &  &  & \checkmark & $0.751\pm 0.125$ & $0.716\pm 0.174$ \\
\midrule
\checkmark & \checkmark &  &  & $0.772\pm 0.129$ & $0.706\pm 0.197$ \\
\checkmark &  & \checkmark &  & $0.766\pm 0.131$ & $0.740\pm 0.181$ \\
\checkmark &  &  & \checkmark & $0.778\pm 0.136$ & $0.711\pm 0.199$ \\
 & \checkmark & \checkmark &  & $0.760\pm 0.134$ & $0.687\pm 0.246$ \\
 & \checkmark &  & \checkmark & $0.758\pm 0.138$ & $0.694\pm 0.226$ \\
 &  & \checkmark & \checkmark & $0.761\pm 0.134$ & $0.711\pm 0.212$ \\
\midrule
\checkmark & \checkmark & \checkmark &  & $0.765\pm 0.132$ & $0.731\pm 0.190$ \\
\checkmark & \checkmark &  & \checkmark & $0.773\pm 0.137$ & $0.716\pm 0.192$ \\
\checkmark &  & \checkmark & \checkmark & $0.766\pm 0.132$ & $0.740\pm 0.184$ \\
 & \checkmark & \checkmark & \checkmark & $0.760\pm 0.135$ & $0.706\pm 0.219$ \\
\midrule
\checkmark & \checkmark & \checkmark & \checkmark & $0.768\pm 0.132$ & $0.740\pm 0.181$ \\
\bottomrule
\end{tabular}
\end{table}

\textbf{Two complementary properties owned by different terms:} Ranking-driven decisions (top-$r$ source selection, $k$-medoids subset selection) depend only on relative orderings (Spearman). Soft-selection or distance-weighted procedures additionally need approximate linear $\mathcal{D}\approx aE+b$ (Pearson).

\textbf{The unrefined baseline establishes a non-trivial starting point with high variance.} Pearson $0.725$, Spearman $0.653$, largest run-to-run variance. The geometry reflects generic LWM proximity rather than the functional notion of cross-dataset transfer.

\textbf{Listwise is the dominant single source of Pearson gain.} Switching only listwise on raises Pearson to $0.777$ (largest single-term gain) and cuts variance by $\approx 30\%$. The loss directly penalizes disagreement between the ranking induced by similarities and the ranking implied by $\mathbf{P}$ over all sources for a fixed target, which simultaneously lifts ranking and linear agreement.

\textbf{Distillation produces the largest single-term Spearman.} Distillation alone reaches Spearman $0.716$ while leaving Pearson modest ($0.751$). It encourages nearest-neighbor relations between the SW teacher and the centroid student, improving rank consistency but not by itself guaranteeing a calibrated linear relation.

\textbf{Consistency and correlation are variance reducers.} Consistency-only ($0.759$) and correlation-only ($0.762$) leave the geometry close to baseline. Consistency averages stochastic-view noise; correlation is a global affine rescaling that cannot fix ordering violations on its own. Both become useful in combination with listwise: listwise+distillation reaches Pearson $0.778$/Spearman $0.711$; listwise+correlation reaches $0.766/0.740$. The full four-term objective ($0.768/0.740$) is essentially the envelope of the best two-term combinations, explaining the diminishing returns of adding more terms. The decomposition is therefore \emph{ordering} (listwise) $+$ \emph{neighborhood shape} (distillation) $+$ \emph{calibration/variance} (correlation, consistency), with the first two load-bearing.

\subsection{Distance Metric: Euclidean vs.\ Sliced Wasserstein}
\label{sec:euc_vs_swd}

\begin{figure}[t]
\centering
\begin{subfigure}[b]{0.31\columnwidth}\centering
\includegraphics[width=\linewidth]{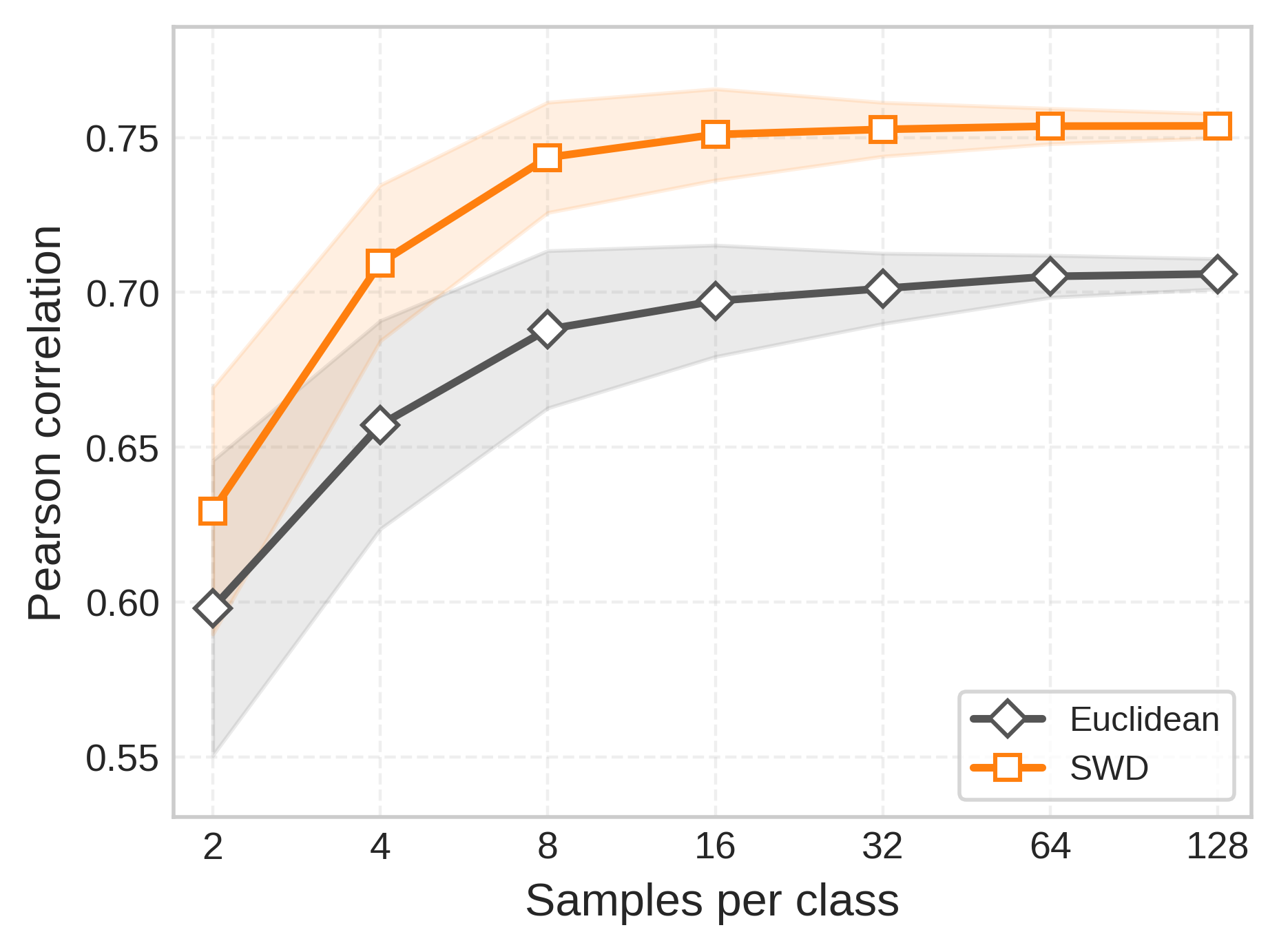}
\caption{Beam: Pearson}\end{subfigure}\hfill
\begin{subfigure}[b]{0.31\columnwidth}\centering
\includegraphics[width=\linewidth]{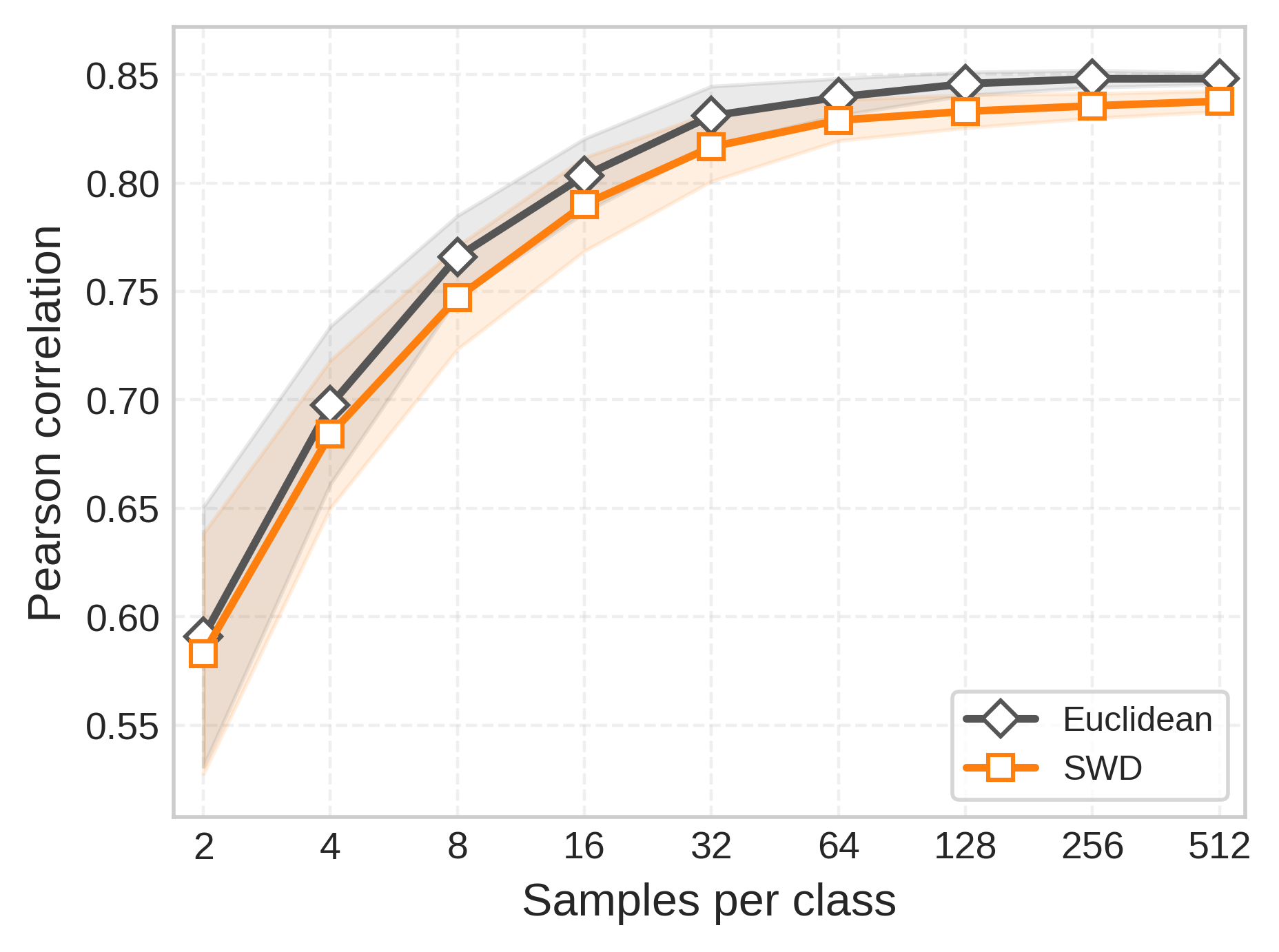}
\caption{LoS: Pearson}\end{subfigure}\hfill
\begin{subfigure}[b]{0.31\columnwidth}\centering
\includegraphics[width=\linewidth]{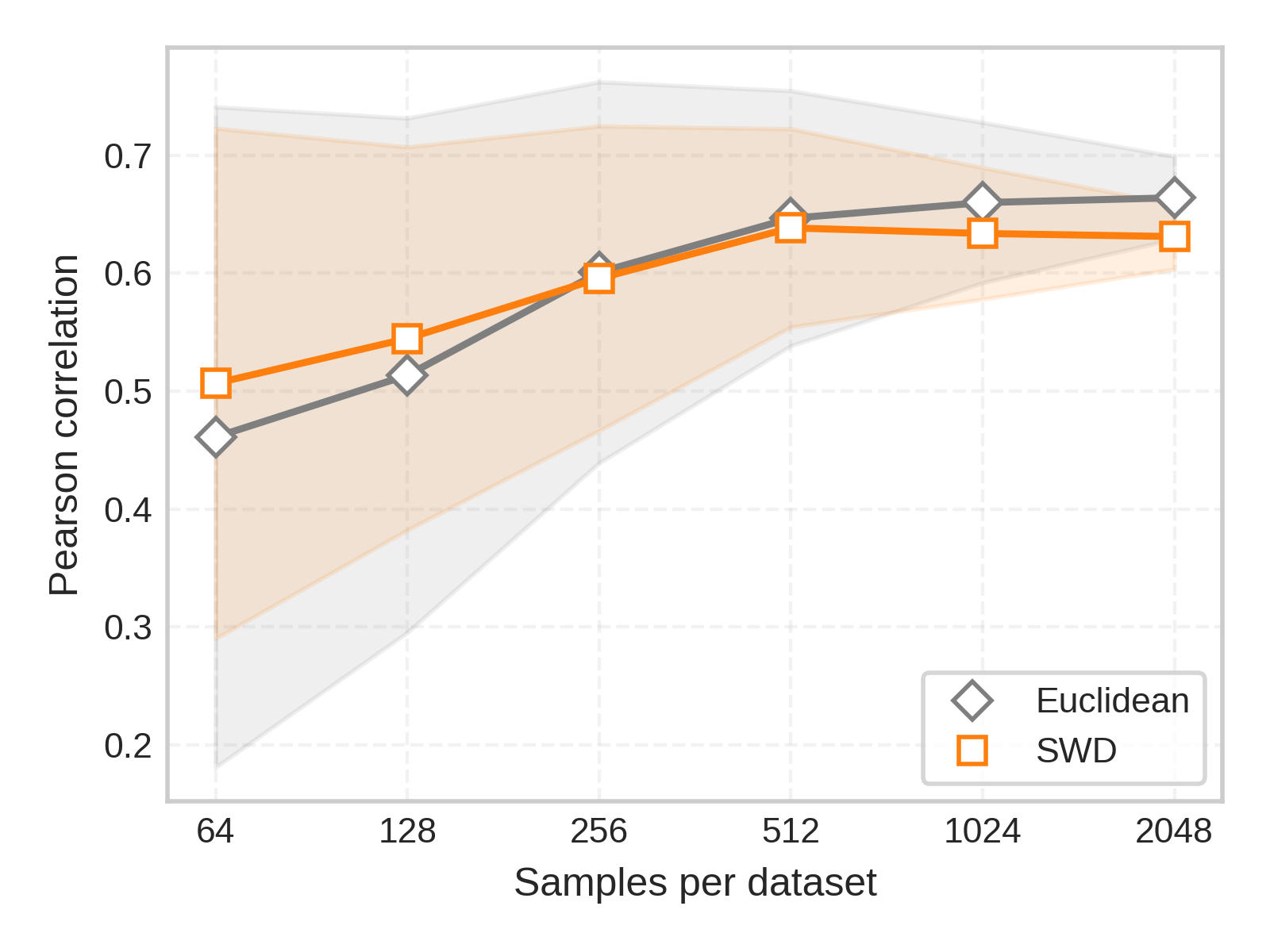}
\caption{CSI compr: Pearson}\end{subfigure}\\[0.4em]
\begin{subfigure}[b]{0.31\columnwidth}\centering
\includegraphics[width=\linewidth]{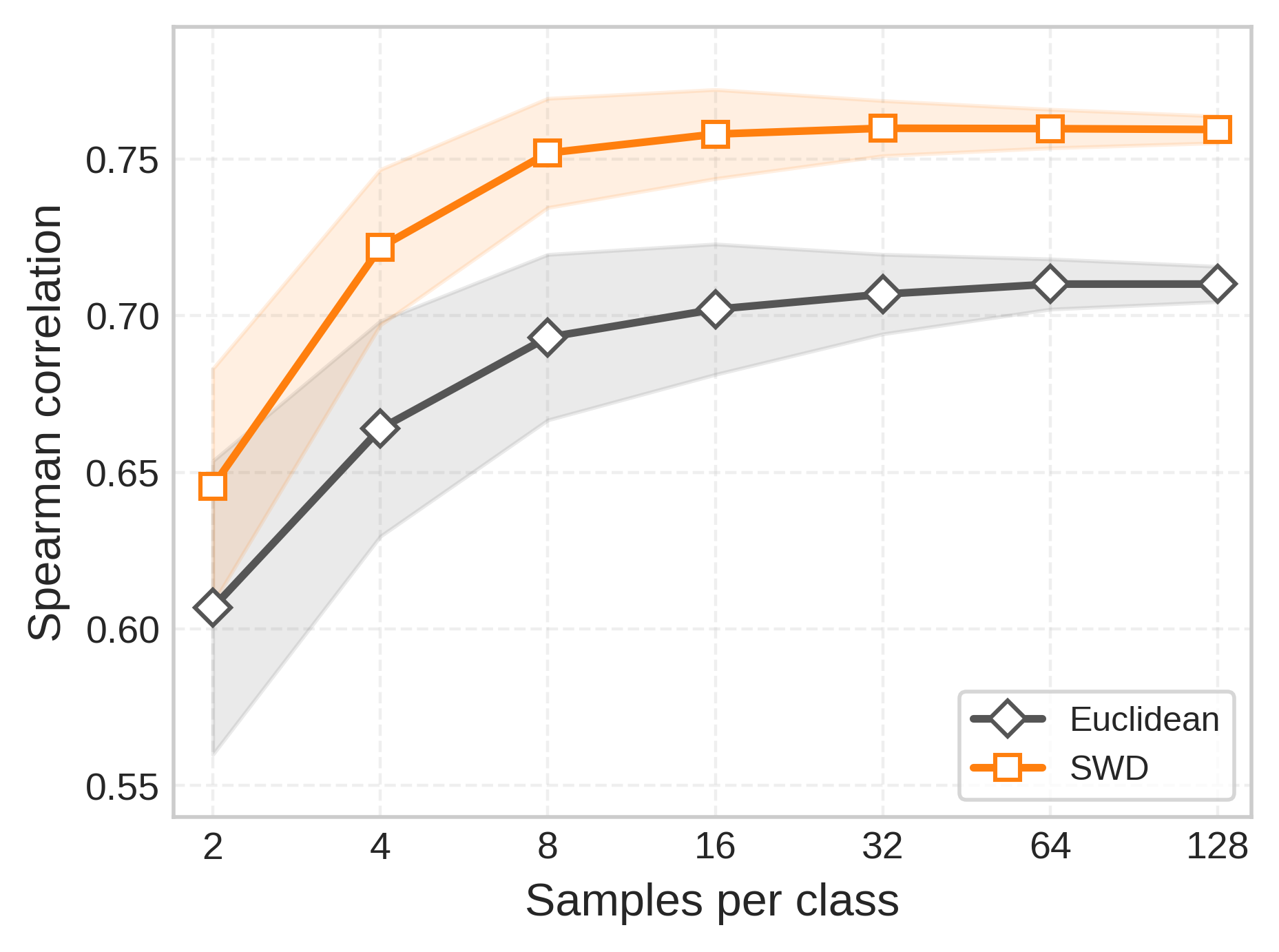}
\caption{Beam: Spearman}\end{subfigure}\hfill
\begin{subfigure}[b]{0.31\columnwidth}\centering
\includegraphics[width=\linewidth]{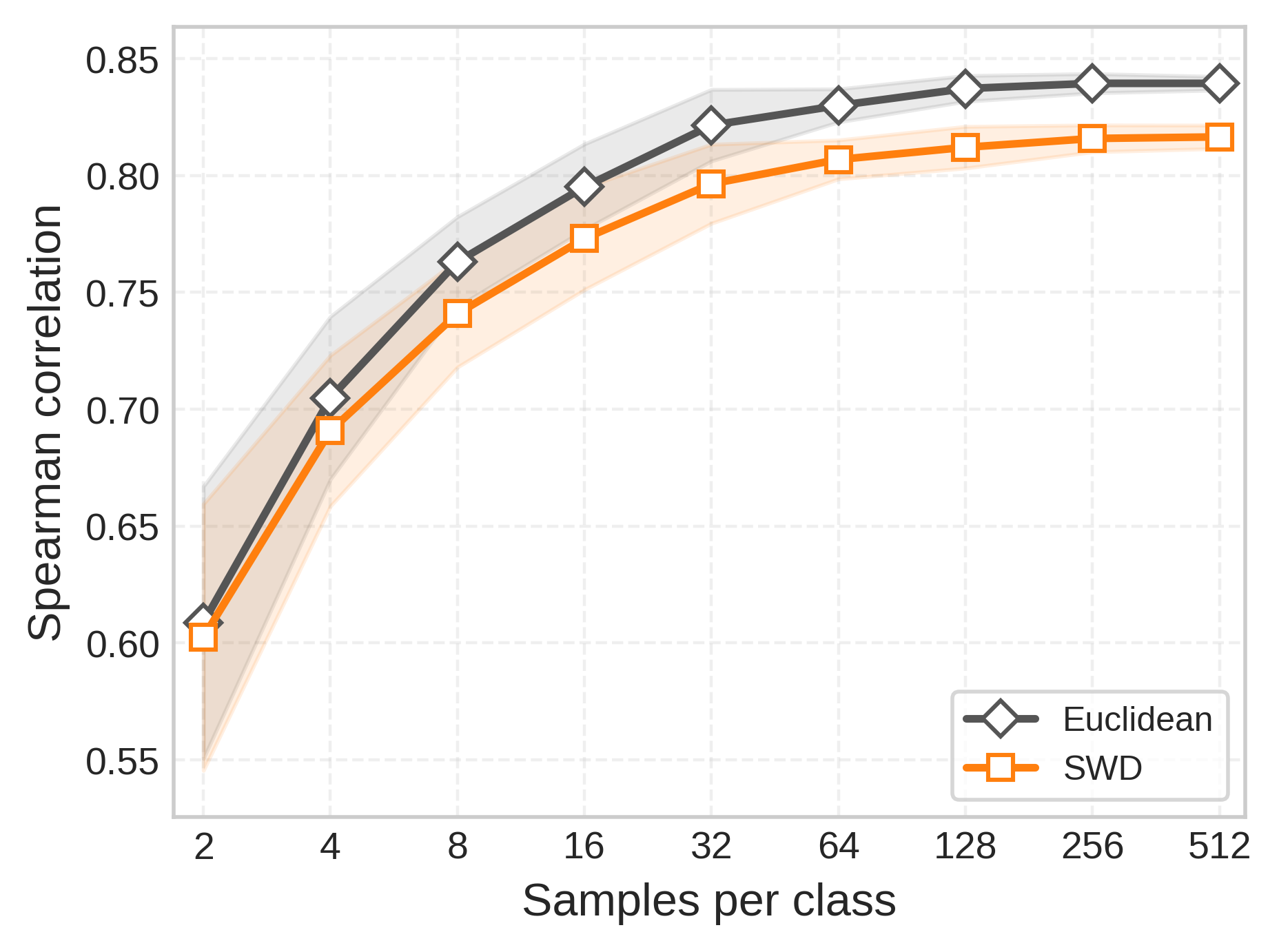}
\caption{LoS: Spearman}\end{subfigure}\hfill
\begin{subfigure}[b]{0.31\columnwidth}\centering
\includegraphics[width=\linewidth]{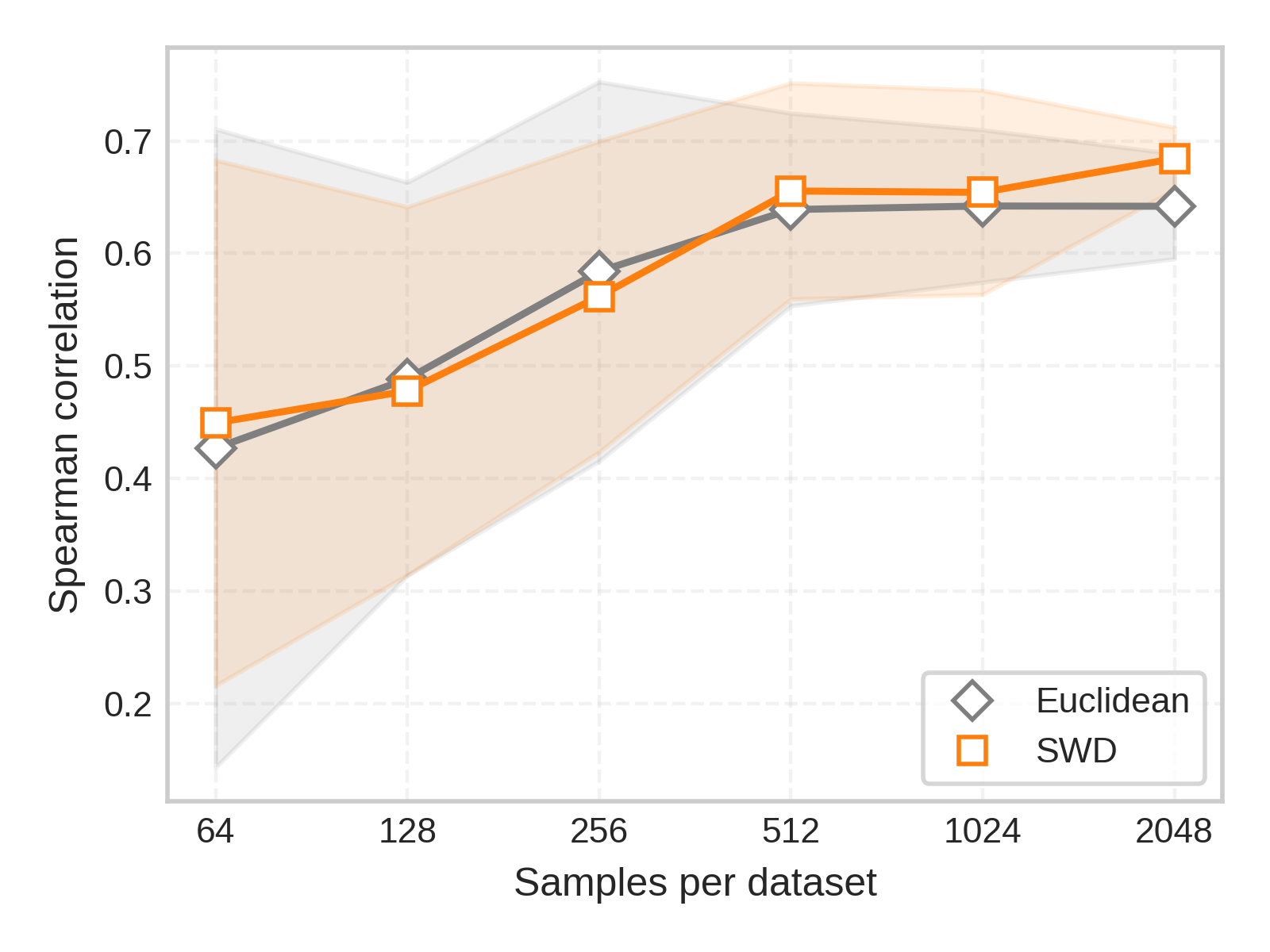}
\caption{CSI compr: Spearman}\end{subfigure}
\caption{\textbf{Correlation between dataset distances and transfer performance as a function of dataset size.} Pearson (top) and Spearman (bottom) for beam prediction, LoS/NLoS, and CSI compression as samples per dataset grow.}
\label{fig:euc_wass_sample_size}
\end{figure}

\begin{figure}[t]
\centering
\begin{subfigure}[b]{0.48\linewidth}\centering
\includegraphics[width=\linewidth]{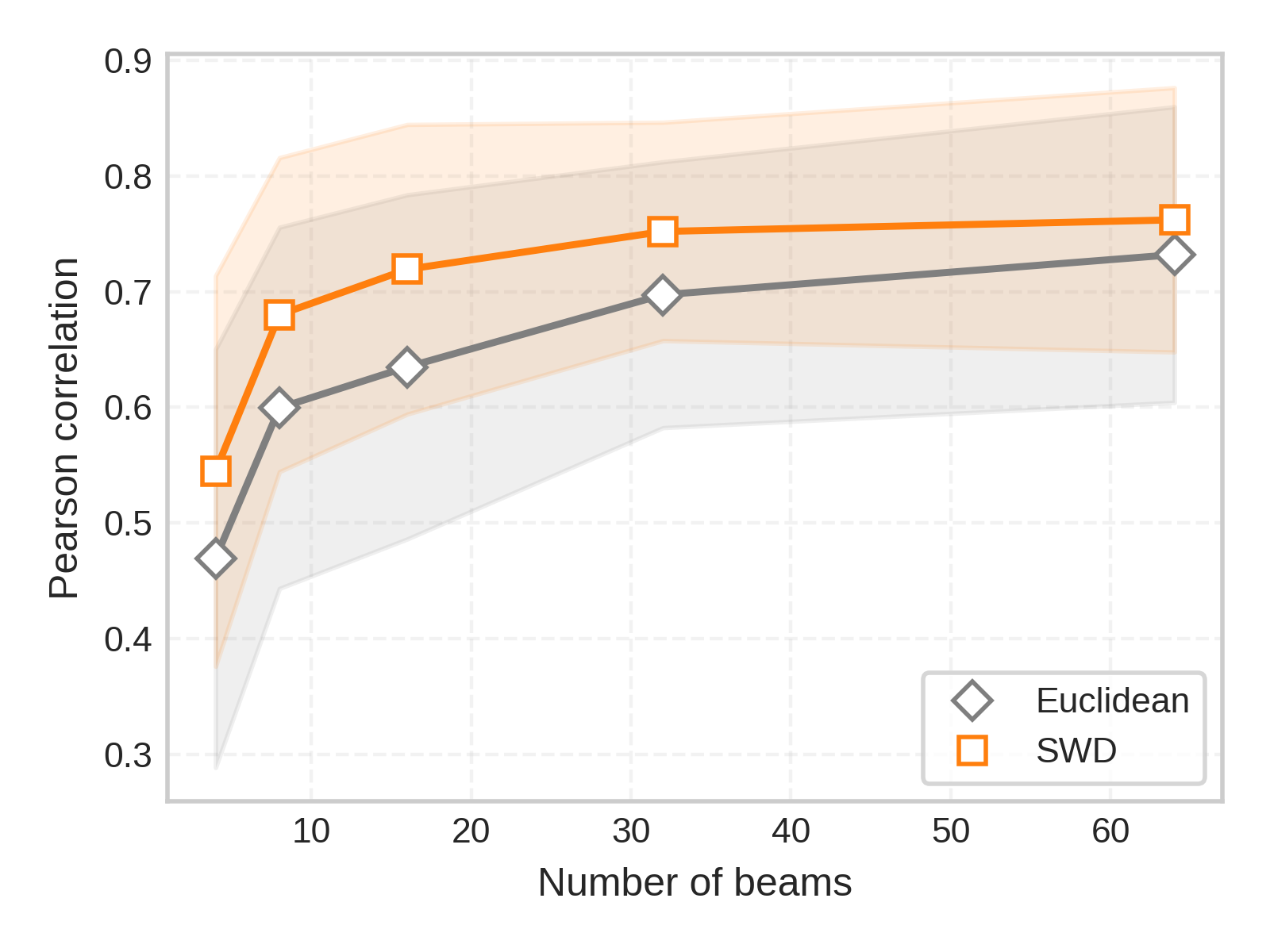}
\caption{Pearson}\end{subfigure}\hfill
\begin{subfigure}[b]{0.48\linewidth}\centering
\includegraphics[width=\linewidth]{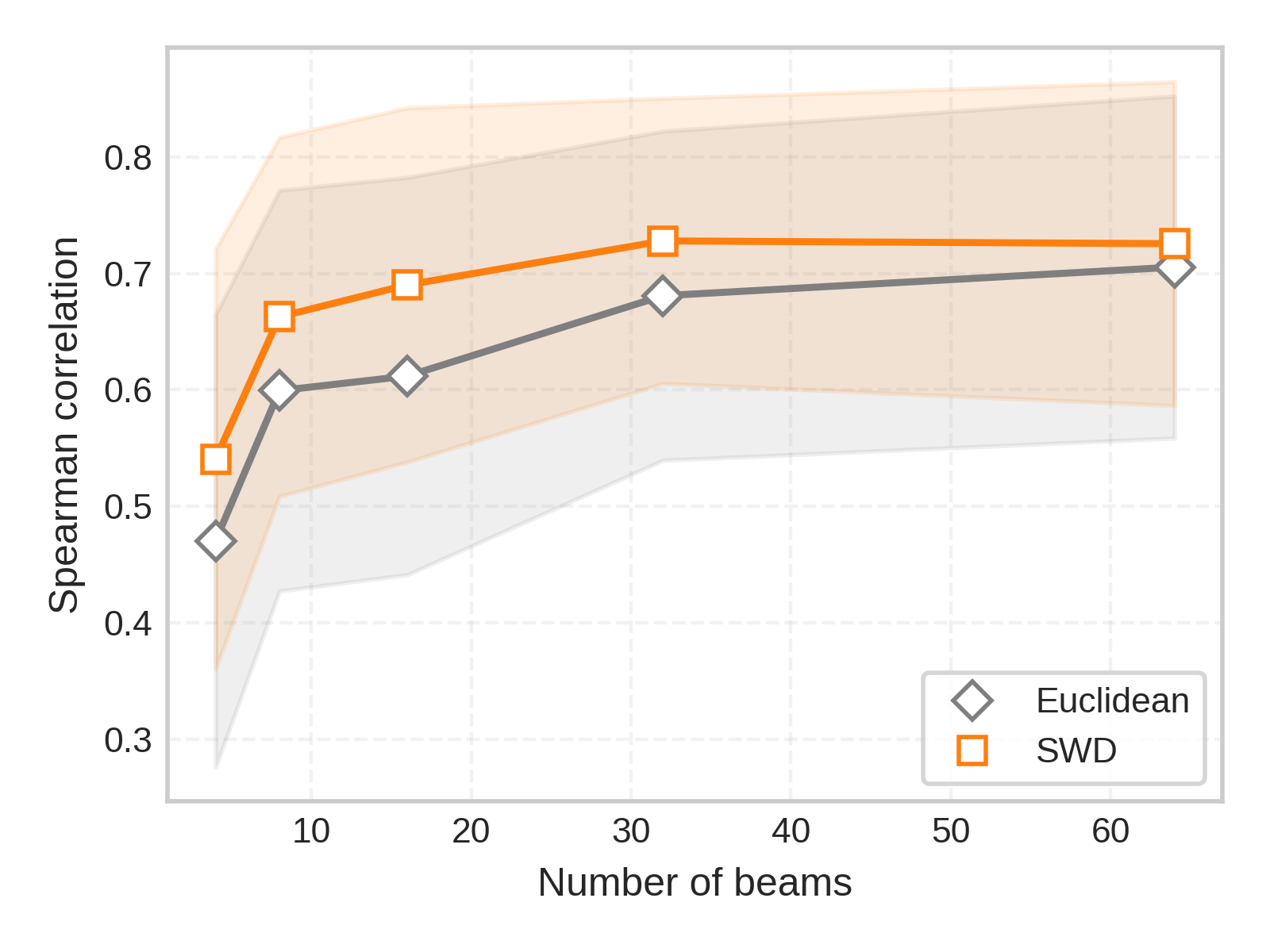}
\caption{Spearman}\end{subfigure}
\caption{Effect of task difficulty on Euclidean and sliced Wasserstein transfer proxies. Beam prediction as the number of beams grows. SW generally achieves stronger correlations across task difficulty, with the gap diminishing as the task becomes more complex.}
\label{fig:euc_wass_complexity}
\end{figure}

\begin{figure}[t]
\centering
\begin{subfigure}[t]{0.235\linewidth}\centering
\includegraphics[width=\linewidth]{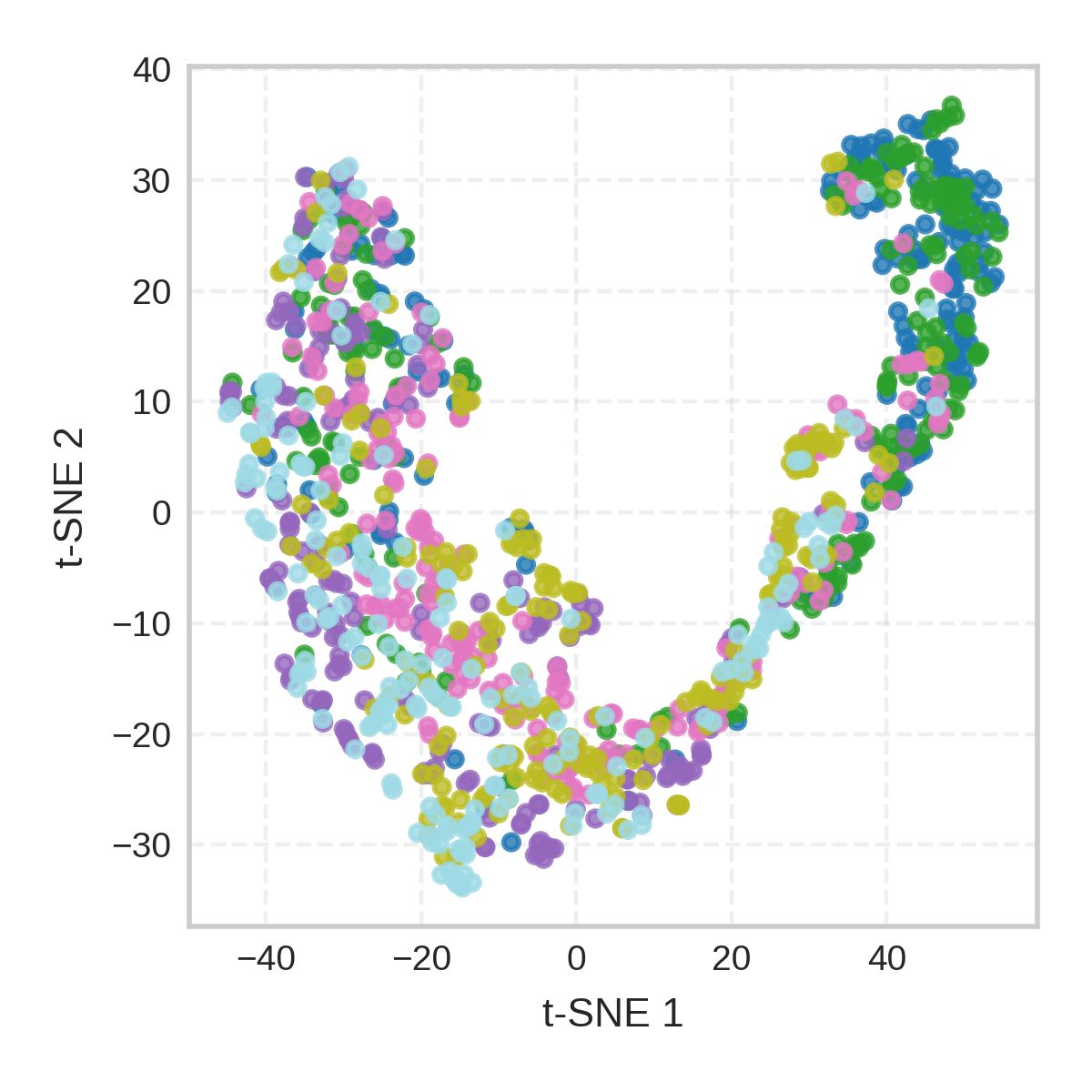}
\caption{Beam: datasets}\end{subfigure}\hfill
\begin{subfigure}[t]{0.235\linewidth}\centering
\includegraphics[width=\linewidth]{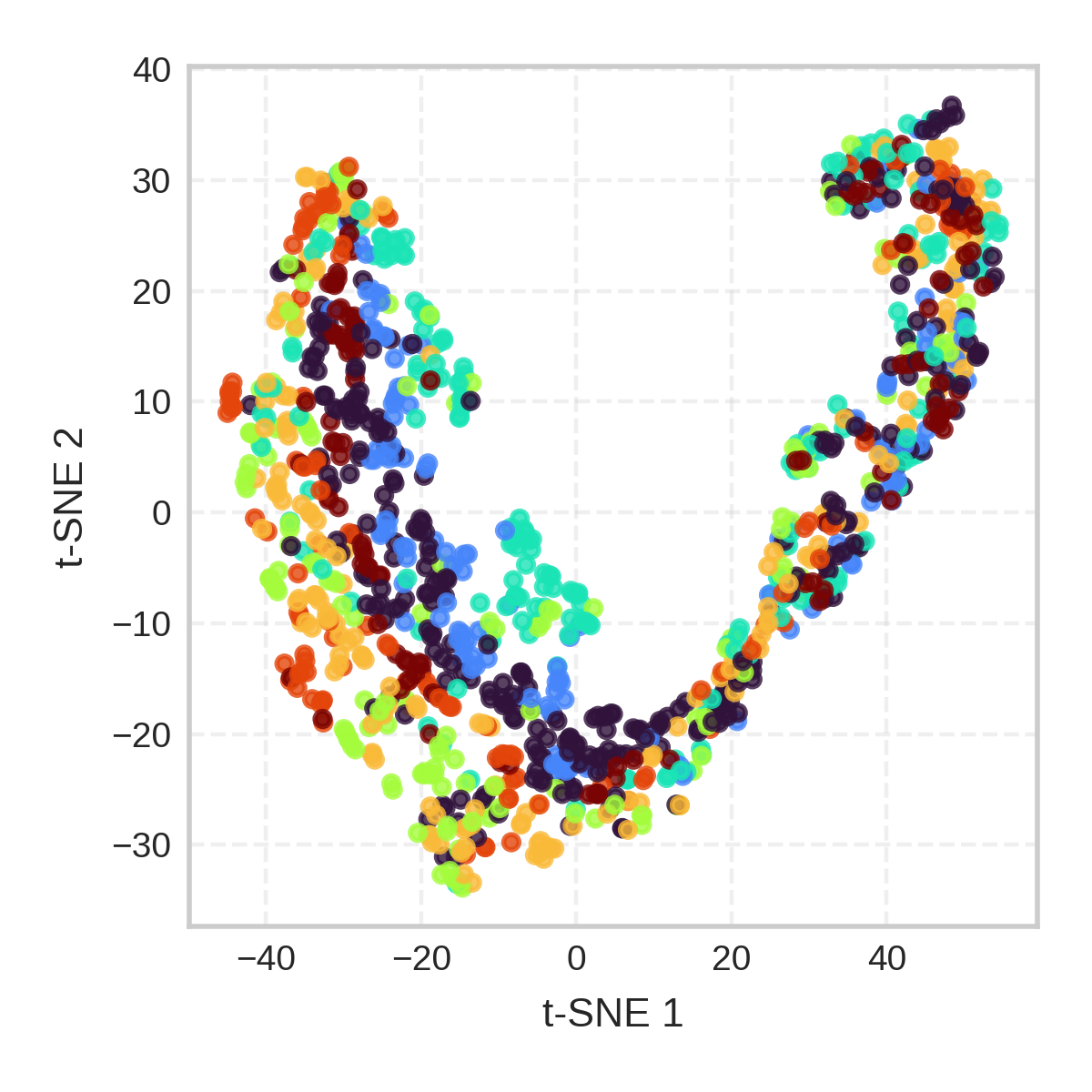}
\caption{Beam: labels}\end{subfigure}\hfill
\begin{subfigure}[t]{0.235\linewidth}\centering
\includegraphics[width=\linewidth]{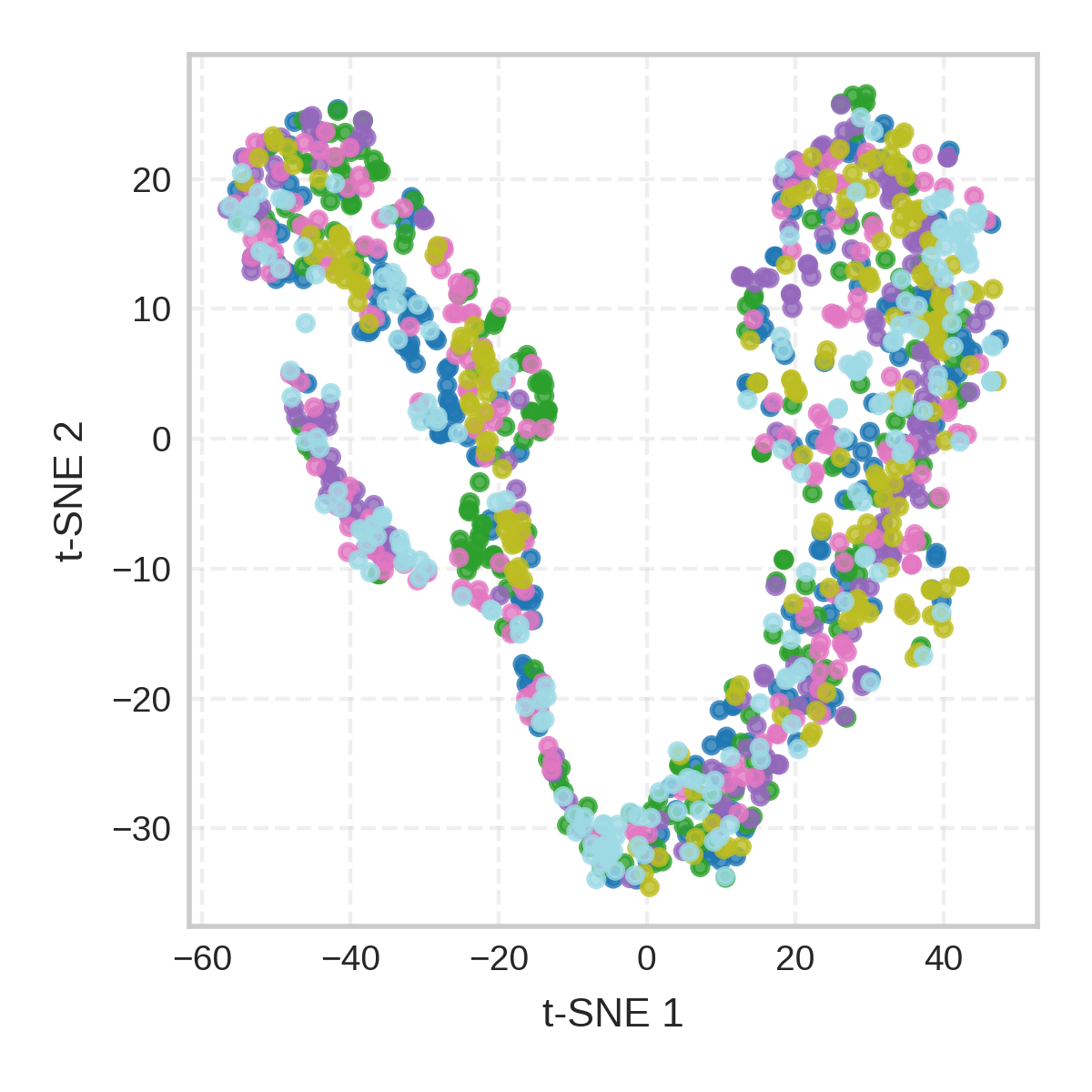}
\caption{LoS: datasets}\end{subfigure}\hfill
\begin{subfigure}[t]{0.235\linewidth}\centering
\includegraphics[width=\linewidth]{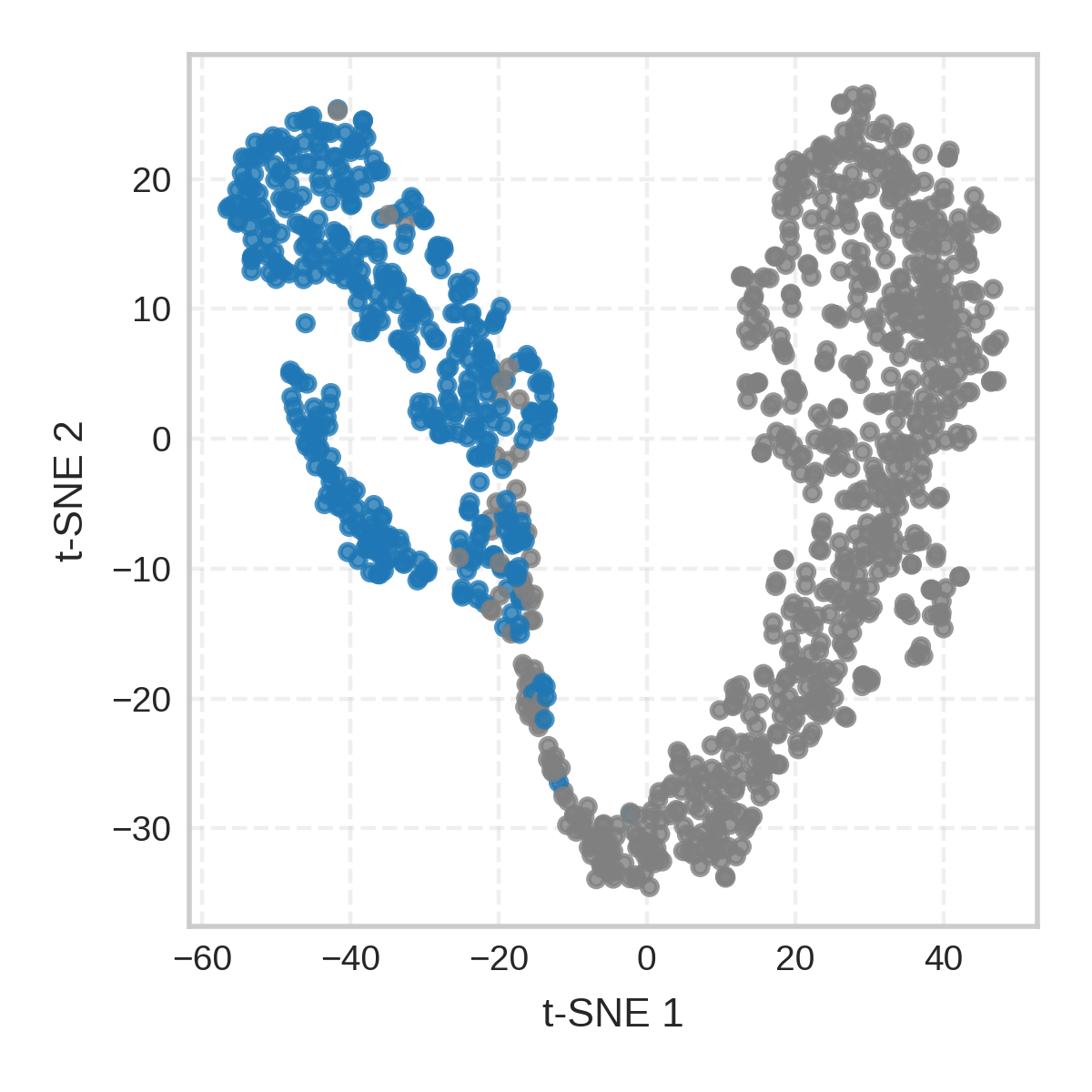}
\caption{LoS: labels}\end{subfigure}
\caption{t-SNE visualization of dataset embeddings, colored by dataset identity (a, c) and by label (b, d), illustrating how the learned representation organizes both dataset structure and task-relevant semantics.}
\label{fig:tsne}
\end{figure}

We compare \emph{label-aware Euclidean centroid distance} and \emph{label-aware sliced Wasserstein distance} as proxies for dataset transfer. Both can be viewed as distances between class-conditional empirical measures: the centroid distance compares first moments $\widehat{m}_i^{(c)}=\frac{1}{m_i^{(c)}}\sum_{z\in Z_i^{(c)}}z$ via $\|\widehat{m}_i^{(c)}-\widehat{m}_j^{(c)}\|_2$, while SW compares the full class-conditional distributions over $L$ Monte Carlo directions $u_\ell\sim\mathcal{U}(\mathbb{S}^{d-1})$. To mitigate class imbalance we use balanced subsampling with $k_c=\min(m_i^{(c)},m_j^{(c)},M)$ per shared class. The Euclidean variant is low-variance and captures only first-moment mismatch; SW captures multi-modality and shape but costs $\mathcal{O}(Lk_cd)+\mathcal{O}(Lk_c\log k_c)$ per pair.

\textbf{Effect of sample size:} \cref{fig:euc_wass_sample_size} reports correlations as samples per dataset grow. Both improve monotonically and saturate, reflecting empirical-estimator convergence; the saturation point differs across tasks because class-conditional distributions have different intrinsic variability. \textbf{For beam prediction, SW achieves stronger correlations across sample sizes}, indicating that capturing higher-order distributional mismatch beyond means is beneficial in this multi-modal multi-class setting. \textbf{For LoS/NLoS, centroid is comparable or slightly better at moderate budgets}: \cref{fig:tsne} confirms that class-conditional clusters are compact and well separated, making centroids low-variance sufficient statistics. For CSI compression (label-free reconstruction), both behave similarly with overlapping uncertainty bands.

\textbf{Effect of task complexity:} \cref{fig:euc_wass_complexity} examines beam prediction as the number of beams grows. Correlations grow with task difficulty because inter-dataset performance gaps become more structured as cardinality grows. SW maintains an advantage at lower beam counts where class overlap and finite-sample effects are pronounced; as the number of beams grows, the SW-vs-centroid gap narrows and both become strong proxies.

\textbf{Takeaway:} Centroid distances provide a strong, stable, computationally efficient baseline when class-conditional structure is compact. Label-aware SW distances are more expressive when class distributions are multi-modal or overlapping, provided finite-sample estimation of higher-order structure remains reliable.

\subsection{Asymmetric Evaluation against Unsymmetrized $\mathbf{P}$}
\label{sec:analysis_asym}
Symmetric distances are necessarily evaluated against $\tilde{\mathbf{P}}_{ij}=\tfrac{1}{2}(P_{ij}+P_{ji})$, so they cannot resolve the off-diagonal signal $\tfrac{1}{2}(P_{ij}-P_{ji})$. The directed extension of \cref{sec:directed} is designed to recover that signal.

\textbf{$\mathrm{DSW}$ predicts unsymmetrized $\mathbf{P}$ with correlations comparable to symmetric SW against $\tilde{\mathbf{P}}$.} On the held-out test split with the CDE backbone, $\mathrm{DSW}$ achieves Pearson/Spearman $(0.77,0.57)$ on LoS/NLoS and $(0.73,0.65)$ on beam prediction over $10$ trials at $n_{\text{train}}=3000$. Predicting unsymmetrized $\mathbf{P}$ is strictly harder (twice the per-cell variance), yet $\mathrm{DSW}$ remains clearly positive. As a sanity check, evaluating the same symmetric SW against unsymmetrized $\mathbf{P}$ yields uniformly lower correlations, isolating the gain to the directed extension rather than the encoder.

\textbf{The spread-ratio penalty is the source of $\mathrm{DSW}$'s directional power but also its finite-sample fragility.} Under reduced budget ($n_{\text{train}}=n_{\text{test}}=400$, $5$ trials), $\mathrm{DSW}$ obtains Pearson $0.64\pm 0.08$ and Spearman $0.49\pm 0.05$ on LoS/NLoS, lower than the large-budget numbers. The cause is mechanical: per-class standard deviations $\sigma_i^{(c)}$ driving the $\log(\sigma_j^{(c)}/\sigma_i^{(c)})$ penalty are $U$-statistic estimates with $\mathcal{O}(1/m_i^{(c)})$ variance, so under-sampled classes inflate the directed component much more than the symmetric centroid term. In small-budget deployments this argues for either shrinking $\alpha$ in \eqref{eq:dswd_full} or falling back to $\mathrm{DSW}_\pi$ alone.

\textbf{How the per-class budget propagates into the directed penalty:} The previous paragraph attributes $\mathrm{DSW}$'s small-budget drop to estimator variance in $\sigma_i^{(c)}$; here we make the variance budget concrete. For a Gaussian-ish class-conditional embedding, $\widehat{\sigma}_i^{(c)}$ has relative standard error of order $1/\sqrt{2m_i^{(c)}}$, so $\log(\widehat{\sigma}_j^{(c)}/\widehat{\sigma}_i^{(c)})$ inherits a standard deviation of roughly $\sqrt{1/m_j^{(c)}+1/m_i^{(c)}}$ on top of its true value. At $m_i^{(c)}=50$ (the regime an under-represented class in a LoS-skewed source falls into at $n_{\text{train}}=400$) the noise floor on the log-ratio is $\approx 0.2$, comparable to the typical true log-ratio between source$\to$target pairs in our pool; at $m_i^{(c)}=200$ it drops to $\approx 0.1$. The directed component is therefore operating near $1{:}1$ signal-to-noise on its key term at small budgets, even before label corruption. The symmetric-vs-directed crossover is therefore not a property of the LWM space but of the plug-in estimator: an empirical-Bayes shrinkage of $\sigma_i^{(c)}$ toward a per-source pooled scale, or an end-to-end-learned directional component, would push the crossover to smaller $m_i^{(c)}$, which is the design space our companion follow-up will explore.

\textbf{Why we report symmetric distances despite an asymmetric transfer matrix:} Because $\mathbf{P}$ is itself directional, the \emph{correct} distance functional for the problem is the directed one: an ideal deployment would always use $\mathrm{DSW}$ in \eqref{eq:dswd_full} and evaluate it against unsymmetrized $\mathbf{P}$. The choice of which distance to use is not a free design parameter; it is dictated by the asymmetry of the underlying performance signal. The reason we adopt the symmetric distance throughout the main-result protocols is therefore not a methodological preference but a finite-sample fact: in the operator-grade label-quality regimes our experiments span, the spread-ratio penalty that gives $\mathrm{DSW}$ its directional power becomes statistically fragile (label corruption attacks the per-class $\sigma_i^{(c)}$ estimates twice, see also \cref{sec:robustness}). Under these conditions a symmetric distance against $\tilde{\mathbf{P}}$ trades a known information loss (the off-diagonal $\tfrac{1}{2}(P_{ij}-P_{ji})$ component) for stability and comparability with prior work~\cite{morais2024datasetsimilarityevaluationframework,morais2026wirelessdatasetsimilaritymeasuring}, which is the favourable trade in the present setting. Resolving this, that is, designing a learnable asymmetric estimator whose directional sensitivity does not depend on plug-in $\sigma_i^{(c)}$ estimates and therefore degrades gracefully under small per-class budgets, is the subject of a dedicated companion paper to this work (\cref{sec:discussion}).

\subsection{Robustness to Label Corruption}
\label{sec:robustness}
Operator-grade libraries rarely come with clean deterministic labels; labels may be incomplete, automatically generated, or absent. We perturb the labels used in distance computation while keeping $\mathbf{P}$ computed with clean labels: \emph{label noise} flips each label uniformly with probability $\eta\in\{0.20,0.50\}$; \emph{label drop} removes a fraction $\rho\in\{0.20,0.50\}$ and computes label-aware SW on what remains; \emph{no labels} replaces label-aware SW with its label-agnostic counterpart. The setup uses LWM-CDE on LoS/NLoS (strongest native class imbalance) and averages over $10$ trials at $n_{\text{train}}=n_{\text{test}}=400$.

\begin{table}[t]
\centering
\caption{Robustness to label corruption (LoS/NLoS, frozen LWM, $10$ trials). Perturbations are applied \emph{only to the distance computation}; $\mathbf{P}$ uses clean labels throughout.}
\label{tab:robustness}
\setlength{\tabcolsep}{3pt}\renewcommand{\arraystretch}{1.1}
\footnotesize
\begin{tabular}{lccccc|c}
\toprule
& Clean & \multicolumn{2}{c}{Noise} & \multicolumn{2}{c|}{Drop} & No labels \\
& & $20\%$ & $50\%$ & $20\%$ & $50\%$ & \\
\midrule
Pearson  & $.59{\pm}.11$ & $.55{\pm}.14$ & $.51{\pm}.13$ & $.61{\pm}.10$ & $.57{\pm}.11$ & $.49{\pm}.12$ \\
Spearman & $.61{\pm}.08$ & $.56{\pm}.11$ & $.48{\pm}.12$ & $.63{\pm}.08$ & $.60{\pm}.09$ & $.44{\pm}.13$ \\
\bottomrule
\end{tabular}
\end{table}

\textbf{Label noise degrades correlation gradually.} Pearson decays smoothly from $0.59$ (clean) to $0.55$ ($20\%$ noise) to $0.51$ ($50\%$ noise). A flip of fraction $\eta$ contaminates each class-conditional measure with mass from the wrong class, biasing the centroid by an amount proportional to the inter-class separation. The graceful degradation is consistent with the LWM space being well separated between classes.

\textbf{Label drop is asymptotically benign and at moderate rates slightly beneficial.} At $20\%$ drop, Pearson improves marginally to $0.61$; at $50\%$ drop it is $0.57$, statistically indistinguishable from clean. Random removal disproportionately thins the over-represented majority class, giving more balanced per-class supports. Unlike noise, dropping does not introduce cross-class contamination, so the only effect is a sample-size reduction the $\mathcal{O}(1/m^{(c)}_i)$ variance bound absorbs.

\textbf{Operating without any labels degrades correlation but does not collapse it.} Pearson $0.49$ and Spearman $0.44$ in the label-agnostic regime: even without supervision, the LWM space retains enough distributional structure to be predictive of transfer. Label-aware aggregation is a multiplier on an already-useful representation, not a load-bearing assumption.

\textbf{Directed evaluation amplifies the noise and no-labels penalties.} Under $\mathrm{DSW}$, the ordering noise $>$ drop $\approx$ clean is preserved but the gradient is steeper: Spearman at $50\%$ noise drops to $0.23$, and no-labels Pearson collapses to $0.26$. Both the source-prior weighting $\pi_i(c)$ and the spread-ratio penalty depend on class assignments, so corrupting labels attacks $\mathrm{DSW}$ twice while attacking the symmetric component once. This sharper degradation is the direct empirical motivation for our choice to report symmetric distances throughout the main-result protocols (\cref{sec:analysis_asym}): the asymmetric distance is the correct functional in principle, but in finite-sample label-corruption regimes the symmetric variant absorbs label noise more gracefully, so it is the operationally reliable surrogate to use today. Closing this gap with an estimator-grade directed distance is left to follow-up work.

\subsection{Cross-Band Generalization}
\label{sec:xband}
Operator deployments rarely stay inside the band a refinement was trained on. We construct a sub-6\,GHz training pool of $24$ DeepMIMO datasets ($8$ diverse cities $\times 3$ BS), compute $\mathbf{P}_{\text{sub-6}}$, and fine-tune LWM-CDE on this pool for $25$ epochs via \eqref{eq:l_cde}. We separately construct a $10$-dataset $28$\,GHz mmWave pool with its own $\mathbf{P}_{\text{mmW}}$ and evaluate distance--transfer correlation on it with both encoders.

\begin{table}[t]
\centering
\caption{Cross-band geometry: LWM-CDE fine-tuned on sub-6\,GHz, evaluated on held-out mmWave (28\,GHz). The CDE in-distribution row is a sanity check.}
\label{tab:xband}
\setlength{\tabcolsep}{6pt}\renewcommand{\arraystretch}{1.05}
\begin{tabular}{llcc}
\toprule
Encoder & Eval pool & Pearson & Spearman \\
\midrule
Frozen LWM      & mmWave (held-out band) & $0.141$ & $0.291$ \\
LWM-CDE (sub-6) & mmWave (held-out band) & $0.011$ & $0.198$ \\
\midrule
LWM-CDE (sub-6) & sub-6 (in-distribution) & $0.659$ & $0.629$ \\
\bottomrule
\end{tabular}
\end{table}

\textbf{CDE does not transfer cross-band; the metric head over-fits band-specific structure.} Frozen LWM retains a weak positive Spearman on mmWave ($0.291$), consistent with the LWM pretraining corpus including both bands. The CDE-refined encoder collapses to Pearson $0.011$ on the same pool, while reaching $0.659$ on the sub-6 pool it was trained on. The mechanism is band-specific: at sub-6, LoS/NLoS dominates cross-source variation and the listwise teacher $q_{ij}\propto\exp(-P_{ij}/\tau)$ trains the metric head to align similarities with the LoS-ratio axis; at 28\,GHz that axis compresses under severe blockage and angular-spread geometry takes over, so the head latches onto the wrong axis. This is a generic failure of any listwise refinement trained against a single $\mathbf{P}$, with a clear deployment rule: maintain one CDE checkpoint per band, or learn a band-conditioned head with an explicit band-identity token.

\subsection{Computational Cost}
\label{sec:cost}
The total wall-clock for $\binom{N}{2}$ dataset distances decomposes as $T_{\text{total}}=T_{\text{embed}}+T_{\text{dist}}$. The qualitative gap between representations is explained almost entirely by the scaling of $T_{\text{embed}}$.

\begin{figure}[t]
\centering
\includegraphics[width=\linewidth]{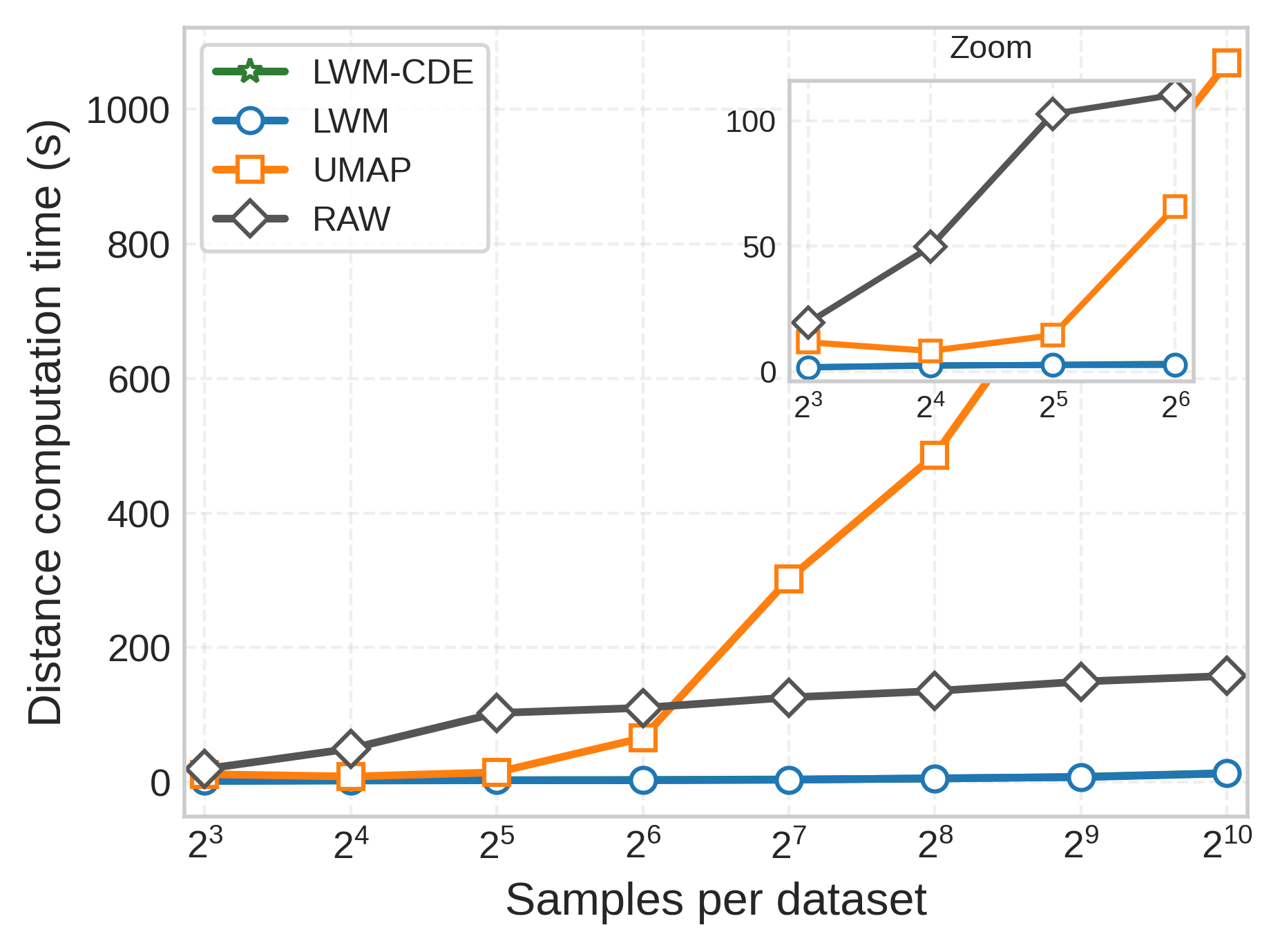}
\caption{Distance computation time using different number of samples per dataset (33 datasets in total).}
\label{fig:distance_delay}
\end{figure}

\textbf{LWM and LWM-CDE scale linearly in the sample budget.} With $t_\phi$ per-sample inference time, $T_{\text{embed}}^{\text{LWM}}\approx Nn\,t_\phi$ is $\mathcal{O}(Nn)$ and independent of the $\binom{N}{2}$ pair count: each embedding is computed once and reused across pairs.

\textbf{UMAP cannot match this scaling} because it is a collection-specific optimization rather than a per-sample encoder. UMAP first builds an approximate $k$-NN graph over all $Nn$ samples and then runs iterative layout optimization with $I$ epochs over $\mathcal{O}(kNn)$ graph edges; the entire procedure must be re-run whenever the collection or sample budget changes.

\textbf{Raw-channel distances suffer doubly}: $d=2MS=2048$ for $32\times 32$ channels inflates the projection term in the SW cost $\mathcal{O}(Lnd)+\mathcal{O}(Ln\log n)$, and they cannot rely on the $\ell_2$-normalization the LWM space provides for free. \cref{fig:distance_delay} confirms the scaling: LWM/LWM-CDE produce essentially flat runtime curves; UMAP is competitive at very small $n$ but scales superlinearly; raw-space distances are the most expensive overall. The one-time CDE fine-tuning cost ($\sim 10$ minutes on a single GPU for our 33-dataset split) is amortized across every subsequent dataset-level decision.

\section{Discussion and Open Problems}
\label{sec:discussion}
The latent space of a wireless foundation model is a reusable \emph{geometric primitive} for the AI-native 6G model lifecycle: frozen LWM embeddings already correlate strongly with empirical transfer (\cref{tab:distancing-compact,tab:dichasus}) orders of magnitude faster than UMAP and raw-channel baselines (\cref{sec:cost}), and LWM-CDE tightens that alignment further when transfer signals are available (\cref{sec:frozen_vs_cde}). The framework lifts any per-sample encoder $\phi:\mathcal{X}\!\to\!\mathbb{R}^d$ to a dataset-level geometry via class-conditional centroid and SW statistics, so it extends naturally to the wider family of wireless foundation models~\cite{alikhani2026lwmtemporal,liu2025wifo,yang2025wirelessgpt,kim2026lwmspectro}. The most consequential extension is to time: two libraries can be indistinguishable under any snapshot-level summary yet transfer differently for time-sensitive tasks (predictive beam tracking, mobility-aware handover), because the difference lives in the joint distribution over consecutive snapshots. A spatiotemporal encoder such as LWM-Temporal~\cite{alikhani2026lwmtemporal} exposes that axis; the same logic extends to frequency via space-time-frequency backbones (WiFo~\cite{liu2025wifo}, addressing the cross-band failure of \cref{sec:xband}) and to multimodal backbones (WirelessGPT~\cite{yang2025wirelessgpt}, LWM-Spectro~\cite{kim2026lwmspectro}).

\textbf{Open problems:} The directed extension of \cref{sec:directed} is the natural distance functional since $\mathbf{P}$ is asymmetric, but its spread-ratio penalty is finite-sample-fragile under operator-grade label quality (\cref{sec:analysis_asym}). A learnable asymmetric metric, with class-prior and class-spread treated as end-to-end estimator components rather than fixed-form penalties, is the subject of a dedicated companion paper. Three further directions are open: \emph{scale}, where operator-grade libraries with hundreds of datasets make per-cell estimation noise on $\mathbf{P}$ the dominant signal; \emph{continual update}, where $\mathbf{P}$ and the library evolve and only the metric head should be re-fit when a new dataset arrives; and \emph{privacy-preserving distancing}, where the fixed encoder enables differentially-private dataset comparison from \emph{statistics of embeddings} rather than raw samples~\cite{dwork2014algorithmic,abadi2016dpsgd}.

\section{Conclusion}
We study dataset-level reasoning for wireless machine learning, where practitioners must decide which datasets to reuse, combine, or prioritize under distribution shift and resource constraints. Treating datasets as points in a learned geometric space, produced by a pretrained wireless foundation model and refined with contrastive objectives aligned to transfer signals, yields inter-dataset distances that closely track empirical cross-dataset transfer and generalize across tasks. Across source selection, label-efficient augmentation, and budgeted pretraining subset selection, distance-guided strategies consistently outperform raw-feature and UMAP baselines. Learned dataset geometries are thus a scalable, reusable abstraction for data-centric decision-making in settings with heterogeneous data and expensive supervision.

\balance
\bibliographystyle{IEEEtran}

\end{document}